\documentclass[aps,twocolumn,superscriptaddress]{revtex4}
\usepackage{mathrsfs}
\usepackage{hyperref}
\usepackage{amsmath}
\usepackage{subfigure}
\usepackage{datetime}

\usepackage{graphicx}
\usepackage{amsmath}
\usepackage{amstext}
\usepackage{amssymb}
\usepackage{amsfonts}
\usepackage{amsbsy}
\usepackage{dcolumn}
\usepackage{bm}
\usepackage{multirow}
\usepackage{color}
\usepackage{amsfonts,amsmath}
\usepackage{epsfig}
\usepackage{latexsym}
\usepackage{bbm}
\usepackage{amssymb}
\usepackage{graphicx}% Include figure files
\usepackage{dcolumn}% Align table columns on decimal point
\usepackage{bm}% bold math
\begin{document}
\title{Low-Temperature Pseudogap Phenomenon: Precursor of High-$T_c$ Superconductivity}
\author{Yao Ma}
\affiliation{Institute for Advanced Study, Tsinghua University,
 Beijing, 100084, P. R. China}
\author{Peng Ye}
 \affiliation{Perimeter Institute for Theoretical Physics, Waterloo, ON, N2L 2Y5, Canada}
\author{Zheng-Yu Weng}
\affiliation{Institute for Advanced Study, Tsinghua University,
 Beijing, 100084, P. R. China}

\begin{abstract}

In this paper, we try to understand the pseudogap phenomenon observed in the cuprate superconductor through a model study.
Specifically, we explore the so-called low-temperature pseudogap state by turning off the superconducting off diagonal long range order
in an ansatz state for the $t$-$J$ model [New Journal of Physics {\bf 13}, 103039 (2011)]. Besides strong non-Gaussian superconducting fluctuations,
the resulting state also exhibits a systematic pseudogap behavior in both spin and charge degrees of freedom, manifested in the uniform
spin susceptibility, specific heat, non-Drude resistivity, Nernst effect,  as well as the quantum oscillation associated with small Fermi pockets emerging in strong magnetic
fields, etc. These anomalous `normal state' properties are found in qualitative consistency with experimental
measurements in the cuprates. Such a model study establishes an intrinsic connection between the peculiar pseudogap properties and the non-BCS nature of the superconducting ground state. Critical comparison with other approaches to the doped Mott insulator is also made.

\end{abstract}
%\pacs{74.40.Kb, 74.72.-h}

% \date{{\small \today}}
%\date{\currenttime, \today}

\maketitle
\tableofcontents
\section{Introduction}

It has been well established\cite{C. C. Tsuei2000,shen_03,Raman2007,STS2007} experimentally that the high-$T_c$ superconductivity in the cuprates shares the same off diagonal long range order (ODLRO), in a form of electron Cooper pairing, with a conventional $d$-wave BCS superconductor. In the BCS theory, turning off such an ODLRO, say, by increasing temperature or applying a strong magnetic field, will lead to a Landau's Fermi liquid description of the normal state\cite{Gennes}. The issue under debate is what is the `normal state' for a cuprate superconductor.

Experimentally it has been observed that the so-called \textit{pseudogap regime} is always present at temperatures higher than the superconducting transition temperature $T_c$, at least in the underdoped regime\cite{timusk_99,shen_03,LNW_06,handbook}. The pseudogap phenomenon has been regarded as one of the most unique physical properties of cuprate superconductors. Whether the pseudogap phase sets the stage for superconductivity to emerge as its low-temperature instability\cite{pwa_87,pwa-book,bza_87,zhang_88,pwa_03,LNW_06,weng_07,gros_07,Tesanovic,zaanen_09,philips_09} or it simply competes with superconductivity as an independent phase\cite{S. A. Kivelson_book,SO(5),S. Sachdev,S. Chakravarty}, has been a crucial issue.

It has been a very challenging quest to understand the nature of pseudogap physics purely phenomenologically.
Alternatively, it is desirable to lay down a general theoretical framework or organizing principles first and then critically examine the experimental facts.
Once the self-consistency of a microscopic theory is established, its comparison with the rich experimental observations, even at a qualitative level, can either lend strong support for it or simply falsify it.

It has been widely accepted that the cuprate superconductors are doped Mott insulators of strong correlations\cite{pwa_87}. At half-filling, the low-energy spin degrees of freedom are properly described by an antiferromagnetic (AF) Heisenberg model, while the charge degree of freedom is gapped due to the strong on-site Coulomb repulsion. The low-lying charge degree of freedom is introduced by doping holes into the system\cite{pwa_87,pwa-book,LNW_06}. The superconducting state appears at low doping,
where the AF long-range order (AFLRO) gets destroyed by the motion of doped holes. Experimentally,
the Mott gap seems to remain finite and large\cite{Mott gap, Mott gap 1}, which guarantees that the charge carriers are the doped holes, while the majority spins in the background are still neutral. Namely, at least in the underdoped cuprates, the superconductivity occurs in a doped Mott insulator regime\cite{pwa_87,pwa-book,LNW_06}.

Some main issues concerning the ground state physics are as follows. Firstly, how the AFLRO at half-filling gets destroyed by doping; Secondly, how the superconductivity arises at finite doping; Thirdly, how the superconductivity begins to diminish at overdoping. Once these are understood, then the next question is, after the superconducting coherence is destroyed by raising temperature or applying strong magnetic fields, what will be the normal state? In particular, if the pseudogap phase is a natural normal state above the superconducting dome in a doped Mott insulator?

Hence, the pseudogap physics can be used as a direct probe into a hypothesized superconducting ground state. The Gutzwiller-projected BCS ground state or the so-called `plain vanilla' resonating-valence-bond (RVB) state has been previously proposed\cite{pwa_87,pwa_03} to describe the superconductivity in the simplest doped Mott insulator, i.e., the two-dimensional (2D) square lattice $t$-$J$ model. The pseudogap properties based on this ground state ansatz has been intensively studied in the literature\cite{LNW_06,gros_07}.

Recently, a new superconducting ground state ansatz has been proposed\cite{Weng_11} for the same $t$-$J$ model. It is distinct from the aforementioned `plain vanilla' RVB ground state by a two-component RVB order instead of the one-component one. In contrast to the `plain vanilla' RVB state, the new ground state can recover an accurate description of the AFLRO state of the Heisenberg model at half-filling (zero doping) limit, while predicts a different (non-BCS) superconductivity at finite doping\cite{Weng_11}.

Most importantly, the new ansatz state can explicitly keep track of a singular sign structure (altered Fermi statistical signs) of the $t$-$J$ model. Specifically, the conventional fermion signs of non-interacting electron gas are completely removed by strong on-site repulsion in the $t$-$J$ model at half-filling, which is described by the Heisenberg model with the ground state precisely satisfying the so-called Marshll sign rule\cite{marshall_55} that is non-statistical trivial sign structure. Nontrivial sign structure (known as the phase string effect\cite{weng_97,WWZ_08}) only emerges upon doping, but is much sparser without bearing any similarity with the conventional fermion signs of the underlying electrons.

Based on such a new superconducting ground state\cite{Weng_11}, we shall present a theoretical description of the low-temperature pseudogap phase in this paper.

Some general properties unique to the present psuedogap phase are found as follows. The state is characterized by a generalized electron fractionalization. Here the correlated electrons are described as if there are three subsystems: holons for doped holes of charge $+e$; neutral spinons of $S=1/2$; and the backflow spinons accompanying the hopping process of the holons. What is particularly simplifying in such a low-temperature pseudogap phase is that the bosonic holons are always Bose-condensed, while the neutral spinons and backflow spinons remain in two-component RVB pairing. In other words, the three subsystems are respectively all in the ODLRO states of their own respectively.

It is well known that an ODLRO in a condensed matter system breaks a global symmetry, resulting in a generalized `rigidity' and great simplications for a many-body system. For these subsystems, the involved symmetries are actually the associated emerging U(1) gauge degrees of freedom upon fractionalization\cite{Weng_11}. The `Meissner' effects due to these ODLROs then suppress the gauge fluctuations, which otherwise would strongly fluctuate and confine these fractionalized particles back to the electrons\cite{LNW_06}. Thus, the hidden ODLROs of the subsystems protect the fractionalization of the electrons self-consistently.

These hidden ODLROs do not necessarily break the \textit{true} global symmetries of the system in general. As it turns out, the aforementioned nontrivial sign structure or phase string effect dictates the above peculier fractionalization of the electrons and mediate the so-called mutual Chern-Simons gauge interaction between the holons and neutral spinons. In contrast to the U(1) gauge fluctuations associated with the fractionalization, such topological gauge fields cannot be `Higgsed' by the hidden ODRLOs and play a crucial role in distroying the superconducting phase coherence in the pseudogap phase.

The pseudogap behaviors are actually the explicit consequences of these hidden ODLROs. For example, the RVB ordering of the spinons is responsible for the pseudogap properties in the uniform magnetic susceptibility and specific heat capacity over a characteristic temperature $T_0$ set by a renormalized superexchange coupling (which decreases with doping), known as the \emph{upper pseudogap phase} (UPP)\cite{GZC2005}. It is in agreement with the early experiments\cite{D. C. Johnston_1989,W. F. Peck1992,T. Imai_1993,M. Ido1994,L. Krusin-Elbaum,J.L. Tallonb2001,Momono}.

The holon condensation further defines a \emph{lower pseudogap phase} (LPP) at lower temperatures. Once it happens at a finite doping, the true AFLRO stops to develop because a doping-dependent, small gap is induced by the mutual Chern-Simons gauge fields in the spin excitation spectrum.  So the LPP is a `spin liquid' (or short-ranged RVB state). It sets a stage for superconductivity to emerge, but the phase coherence is still disordered by thermally excited spin excitations, again via the mutual Chern-Simons gauge fields. Such an LPP is thus featured by strong non-Gaussian superconducting fluctuations.
The longitudinal resistivity is of non-Drude behavior, which is supplemented with a strong Nernst effect.
All of these are clear at variance with the usual Gaussian fluctuations in a narrow critical regime of a BCS superconductor\cite{Gennes}.

These anomalous pseudogap properties in the LPP are qualitatively consistent with the experimental observations in the cuprates \cite{D. C. Johnston_1989,K. C. Ott1991,T. Imai_1993,W. F. Peck1992,M. Ido1994,J. Corson_1999,J.L. Tallonb2001,Ando2004,XZA2000,WYY2001,N. P. Ong,WYY2006,P. H. Hor}. Finally, the true superconducting instability happens when temperature is sufficiently lower than the spin gap, where the fractionalized spin excitations becomes `confined' via the mutual Chern-Simons gauge fields, and their novel phase disordering effect on the superconducting coherence
gets screened out below $T_c$\cite{MW_10}.

It is noted that the LPP as a spin liquid/vortex liquid state has been previously studied by effective theory approaches\cite{WM_02,WQ_06,QW_07}. The present study further provides a microscopic framework based on the wave function description, which handles the short-range correlations more carefully. The main distinction is an emergent two-component RVB structure in the spin background\cite{Weng_11}. Namely, the spin degrees of freedom are now composed of two: a spin liquid always pinned at half-filling together with the so-called \textit{backflow} spinons describing the hopping effect on the spin background.

Such a two-component RVB structure further predicts another non-superconducting state,
which may be obtained with the turning off the ODLRO of the backflow spinons by strong magnetic fields, say, in the magnetic vortex core region.
This core state is to be called the LPP-II, in which the backflow spinons become charged with coherent Fermi pockets, whose Luttinger volume is commensurate with the doped holes.
It is responsible for a novel quantum oscillation and the Pauli behavior of the spin susceptibility,
and provides a consistent explanation for the experiments\cite{QOS2007,GQZ_10,GQZ_12}.

The rest of the paper is organized as follows. In Sec. \ref{sec:effective}, by starting with a new superconducting ground state ansatz for the doped $t$-$J$ model, we introduce the LPP.  An effective Hamiltonian and the corresponding phase diagram from mean field self-consistent calculation will be presented. In Sec. \ref{PhenomenologyLPP} a self-consistent phenomenology for the LPP will be presented,
based on the microscopic effective theory and a comparison with the experiments in the cuprates will be made. In Sec. \ref{NatureLPP}, a critical comparison of the present approach with the slave-boson approach to the $t$-$J$ model will be made. Finally, Sec. \ref{sec:conclusion} will be devoted to a discussion.
An effective topological field theory description of the LPP known as the compact mutual Chern-Simons gauge theory will be also outlined in Appendix \ref{MCSLPP1}.

\section{Low-Temperature Pseudogap Phase as Precursor of Superconductivity}\label{sec:effective}

As emphasized in the Introduction, in this work we will explore the LPP as a `normal state' with the superconducting ODLRO being turned off.
In other words, it is non-superconducting but is most closely related to the superconducting ground state. To characterize such an LPP microscopically, in the following, we start with a new superconducting ground state ansatz, which has been recently proposed\cite{Weng_11} for
the $t$-$J$ model.

\subsection{Motivation: Superconducting ground state ansatz}
\label{fractionalization}

In a doped Mott insulator, the doubly occupied sites are in a high-energy sector due to a large on-site repulsion $U$, which exists only in
virtual processes to mediate the so-called AF superexchange coupling $J=\frac{4t^2}{U}$ between the nearest neighboring (NN)
spins ($t$ is the bare NN hopping integral). At half-filling, with each lattice site occupied by one electron, the relevant degrees of
freedom are localized neutral spins.  At finite doping, the neutral spins remain at singly occupied sites,
and doped charge carriers are at those sites where the electron numbers deviate from the single occupancy. The simplest model to describe the cuprate system as a doped Mott insulator is the so-called $t$-$J$ model\cite{pwa_87,bza_87,ZRS}
\begin{equation}
H_{t\text{-}J}=-t\sum_{\langle ij\rangle\sigma}\hat{c}^{\dag}_{i\sigma}\hat{c}_{j\sigma}+ \text{h.c.} + J\sum_{\langle ij\rangle}
\left(\mathbf{\hat{S}}_i\cdot\mathbf{\hat{S}}_j-\frac{1}{4}\hat{n}_{i}\hat{n}_{j}\right)
\label{HtJ}
\end{equation}%
where $i$,$j$ represent the sites in a 2D square lattice, `$\text{h.c.}$' represents the Hermitian conjugate of the forward term,
and $\hat{c}_{i\sigma}$ is the annihilation operator of an electron at site $i$ with spin index $\sigma$. $\hat{\mathbf{S}}_i$ and $\hat{n}_{i}$ are the spin and number operators, respectively,
at site $i$. In Eq. (\ref{HtJ}), a no-double-occupancy constraint
must be always satisfied: $\hat{n}_{i}\leq1$ for each site $i$ for the hole-doped case. Defined in this restricted Hilbert space,
$\hat{c}_{i\sigma}$ as a bare hole creation operator is not equivalent to the original electron annihilation operator anymore\cite{pwa-book,LNW_06}.

A superconducting ground state has been recently constructed\cite{Weng_11} based on the hole-doped 2D $t$-$J$ model, which is of the following peculiar form of electron fractionalization
\begin{eqnarray}
|\Phi _{\mathrm{G}}\rangle&\equiv &  C \mathcal{\hat{P}}\left(|\Phi _{h}\rangle \otimes |\Phi_{a}\rangle\otimes |\Phi _{b}\rangle  \right)
\label{scgs-0}
\end{eqnarray}%
with $C$ as the normalization factor.

Here, one does not see the electron creation or annihilation operators directly. Instead, the ground state is composed of three subsystems. The $+e$ charge sector of doped holes is described by
\begin{equation}
|\Phi _{h}\rangle \equiv \sum_{\{l_{h}\}}\varphi
_{h}(l_{1},l_{2},...)h_{l_{1}}^{\dagger }h_{l_{2}}^{\dagger }...|0\rangle
_{h}\ ,  \label{bgs}
\end{equation}%
where the bosonic wave function $\varphi _{h}\simeq\text{ constant}$, which defines
a Bose-condensed \textquoteleft holon\textquoteright \ state with $h_{l}^{\dagger }$ acting on a vacuum $|0\rangle _{h}$.
The motion of doped holes will also generate spin backflows, which are described by `itinerant' fermionic $a$-spinons as
\begin{equation}
|\Phi _{a}\rangle \equiv \exp \left( \sum_{ij}\tilde{g}_{ij}a_{i\downarrow
}^{\dagger }a_{j\uparrow }^{\dagger }\right) |0\rangle _{a}, \label{phia-0}
\end{equation}%
where $a_{i\sigma }^{\dagger }$ acts
on a vacuum $|0\rangle _{a}$ and the $a$-spinons are paired with an RVB amplitude $\tilde{g}_{ij}$.

The main spin background is described by
\begin{equation}
|\Phi _{b}\rangle = \exp \left( \sum_{ij}W_{ij}b_{i\uparrow }^{\dagger
}b_{j\downarrow }^{\dagger }\right) |0\rangle _{b} \label{phirvb}
\end{equation}
in which the neutral \textquoteleft spinons\textquoteright \ are RVB-paired with an amplitude $W_{ij}$ in
Eq. (\ref{phirvb}), where $b_{i\sigma }^{\dagger }$ as a bosonic $b$-spinon creation operator acts on a vacuum $|0\rangle _{b}$.

In the superconducting ground state (\ref{scgs-0}), the no-double-occupancy constraint
$\hat{n}_{i}\leq1$ is enforced by the projection operator
\begin{equation}
\mathcal{\hat{P}}\equiv \hat{P}_{\mathrm{B}}\hat{P}_{s}   ,
 \label{P}
\end{equation}
where $\hat{P}_{s}$ enforces the single-occupancy constraint $\sum_{\sigma} n_{i\sigma }^{b}=1$ ($n_{i\sigma }^{b}\equiv b_{i\sigma }^{\dagger }b_{i\sigma }$). Namely, the state
\begin{equation}
|\mathrm{RVB}\rangle\equiv \hat{P}_s|\Phi _{b}\rangle  ,
\label{lda}
\end{equation}
is a neutral spin background which is always pinned at half-filling. The localized spin background $|\mathrm{RVB} \rangle$ is a Liang-Docout-Anderson (LDA)\cite{lda_88} type of bosonic RVB state.
At half-filling, $|\Psi _{\mathrm{G}}\rangle =C |\mathrm{RVB}\rangle$, and the LDA state can describe the AF ground state very accurately in this limit\cite{lda_88,Weng_11}.

$\hat{P}_{\mathrm{B}}$ in Eq. (\ref{P}) further enforces the constraint that the $h$-holon and $a$-spinon always satisfy
\begin{equation}
\text{ }n_{i\bar{\sigma}}^{a}=n_{i}^{h}n_{i\sigma }^{b}
\label{PB}
\end{equation}%
such that each $a$-spinon always coincides with a holon as $\sum_{\sigma
}n_{i\bar{\sigma}}^{a}=n_{i}^{h}$  (here $n_{i\bar{\sigma}}^{a}\equiv a_{i\bar{\sigma}%
}^{\dagger }a_{i\bar{\sigma}}$ and $n_{i}^{h}\equiv h_{i}^{\dagger }h_{i}$
with $\bar{\sigma}\equiv -\sigma )$. Namely, at the hole site, the $a$-spinon will compensate the neutral $b$-spinon in $|\mathrm{RVB}\rangle$
which is an empty site physically. The $a$-spinons will accompany the hopping of the holons, and play a crucial role in keeping track of the effect of
itinerant motion of the doped holes on the spin degrees of freedom.

In the fractionalized form Eq. (\ref{scgs-0}), there is no trace of the original electrons. In fact, the original electron $\hat{c}$-operator, acting on the ground state (\ref{scgs-0}), can be expressed as follows\cite{Weng_11}
\begin{eqnarray}
\hat{c}_{i\sigma } = \mathcal{\hat{P}} h_{i}^{\dagger }a_{i\bar{\sigma}}^{\dagger}(-\sigma)^ie^{i\hat{\Omega}_i} \mathcal{\hat{P}}~,
\label{decomp}
\end{eqnarray}
where $(-\sigma)^i$ is a staggered sign factor introduced for convenience and the phase shift operator $\hat{\Omega}_{i}$ will sensitively depend on spin correlations in $|\Phi _{b}\rangle$, as defined by\cite{Weng_11}
\begin{equation}
\hat{\Omega}_{i}= \frac{1}{2} \left(\Phi_i^s-\Phi_i^0\right)   ,
 \label{phif}
\end{equation}
where
\begin{equation}
\Phi^s_{i}=  \sum_{l\neq i}\theta _{i}(l)\left( \sum_{\sigma }\sigma
n_{l\sigma }^{b} \right)  \label{phis}
\end{equation}%
and
\begin{equation}
\Phi^0_{i}=\sum_{l\neq i}\theta _{i}(l)
\label{phi0}
\end{equation}
in which $\theta _{i}(l)=\mathrm{Im}\ln $ $(z_{i}-z_{l})$ ($z_{i}$ is the
complex coordinate of site $i$).

Here the $\hat{c}$-operator defined in Eq. (\ref{decomp}) is only a hole-creation operator. The corresponding SU(2) spin operators associated with
doped holes are given by
\begin{equation}
\hat{S}_{i}^{az}\equiv \frac{1}{2}\sum_{\sigma }\sigma a_{i\sigma }^{\dagger
}a_{i\sigma }  \label{saz}
\end{equation}%
and%
\begin{equation}
\hat{S}_{i}^{a+}\equiv -(-1)^{i}a_{i\uparrow }^{\dagger }a_{i\downarrow }\text{, \  \  \  \  \  \ }\hat{S}_{i}^{a-}\equiv - (-1)^{i}a_{i\downarrow
}^{\dagger }a_{i\uparrow } ~. \label{sa+}
\end{equation}%
On the other hand, acting on the neutral spin state in Eq. (\ref{scgs-0}), the SU(2) spin operators for $b$-spinons are given as follows
\begin{equation}
\hat{S}_{i}^{bz}\equiv \frac{1}{2}\sum_{\sigma }\sigma b_{i\sigma }^{\dagger
}b_{i\sigma }  \label{sbz}
\end{equation}%
and%
\begin{equation}
\hat{S}_{i}^{b+}\equiv (-1)^{i}b_{i\uparrow }^{\dagger }b_{i\downarrow }e^{i\Phi^h_i}\text{, \  \  \  \  \  \ }\hat{S}_{i}^{b-}\equiv (-1)^{i}b_{i\downarrow
}^{\dagger }b_{i\uparrow } e^{-i\Phi^h_i}~, \label{sb+}
\end{equation}%
from the original Schwinger-boson representation of spin operators with, say, $\hat{S}_{i}^{b+}\equiv (-1)^{i}b_{i\uparrow }^{\dagger }b_{i\downarrow }$. Here
\begin{equation}
\Phi^h_{i}=  \sum_{l\neq i}\theta _{i}(l)
n_{l }^{h}
\label{phih}\end{equation}%
which entangles the $b$-spinons with the $h$-holons non-locally. Finally, the total local spin operators are given by
\begin{equation}
\hat{\bf{S}}_i=\mathcal{\hat{P}} \left(\hat{\bf{S}}_i^b+\hat{\bf{S}}^a_i\right) \mathcal{\hat{P}}~.
\end{equation}

Hence, in such a fractionalized doped-Mott-insulator superconductor, the fundamental spin and charge degrees of freedom are described by three subsystems,
all possessing ODLROs of their own. The charge carriers as bosons experience a Bose condensation in $|\Phi _{h}\rangle $. The spins are effectively characterized by a
two-fluid state, which shares the similarity with other proposals in different contexts\cite{pines_09,wen_05,KLW_09,YW_13}. Here one type is of local moment character for a Mott insulator: they form bosonic RVB pairing in $|\Phi _{b}\rangle$,
which can recover the correct description of the antiferromagnetism in the half-filling limit\cite{lda_88,Weng_11}, while become a short-range AF (spin liquid) state in the superconducting phase; The other type is of itinerant character associated with doping:
they are fermions forming BCS-like pairing. Such a superconducting state will thus show distinctive `rigidity' associated with different ODLROs hidden in its charge and spin components.

\subsection{$d$-wave superconductivity}
\label{d-waveSC}

\subsubsection{Superconducting order parameter}

Now let us examine the superconducting ODLRO in Eq. (\ref{decomp}). The electron singlet pair operator
\begin{eqnarray}
\hat{\Delta}^{\text{SC}}_{ij}&\equiv& \sum_{\sigma }\sigma \hat{c}_{i\sigma
}\hat{c}_{j\bar{\sigma} } \nonumber \\
&= & \hat{\Delta}_{ij}^{0} F_{ij}
\label{sco}
\end{eqnarray}
in which the second line acts on the fractionalized state (\ref{scgs-0}).

Here the phase part of $\hat{\Delta}^{\text{SC}}_{ij}$ is defined by
\begin{equation}
F_{ij} \equiv  e^{i(1/2)\left[\Phi^s_i(j)+\Phi^s_j(i)\right]} ~.
\label{sco2}
\end{equation}

The amplitude part of the Cooper pairing is given by
\begin{equation}
\hat{\Delta}_{ij}^{0} \equiv \mathcal{\hat{P}} \left(h^{\dagger}_ih^{\dagger}_j\hat{\Delta}_{ij}^ae^{-i\phi^0_{ij}}\right) (-1)^je^{-i\Phi^0_j}\mathcal{\hat{P}} ~,
\label{sco1}
\end{equation}
where the pairing operator of the $a$-spinon is defined by
\begin{equation}
\hat{\Delta}_{ij}^{a}\equiv \sum_{\sigma }\sigma a_{i\sigma
}^{\dagger }a_{j\bar{\sigma} }^{\dagger }   .
\end{equation}
And $\phi^0_{ij}$ is a non-dynamic $\pi$ flux per plaquette in a square lattice which is defined as
\begin{equation}
\phi^0_{ij}\equiv\frac{1}{2}\sum_{l\neq i,j}\left[\theta_i(l)-\theta_j(l)\right]   .
\end{equation}
In the following, for simplicity, we shall focus on the case of the NN sites, i.e., $i=NN(j)$, in examining
the superconducting order parameter. In the ground state of Eq. (\ref{scgs-0}), the holon condensation and $a$-spinon pairing
$\langle\hat{\Delta}_{ij}^a e^{-i\phi^0_{ij}}\rangle\neq 0$ will lead to an $s$-wave $\langle\hat{\Delta}_{ij}^0\rangle= \text{constant}$. The phase factor $(-1)^je^{-i\Phi^0_j}$ is a constant independent of site index j, which may be easily shown by noting that
$(-1)^je^{-i\Phi^0_j}\times \left[(-1)^ie^{-i\Phi^0_i}\right]^*=e^{i2\phi^0_{ij}}=1$ for $i=NN(j)$ with a proper gauge choice of
$\phi_{ij}^0$ in evaluating $\langle\hat{\Delta}_{ij}^a e^{-i\phi^0_{ij}}\rangle$ (see below).

The superconducting phase coherence will be determined by
\begin{eqnarray}
\langle F_{ij} \rangle \neq 0 ~.
\label{sco2}
\end{eqnarray}
Note that here $\Phi^s_i(j)$ is different from
$\Phi^s_i$ defined in Eq. (\ref{phis}) by that $l=j$ should be removed in the summation over site $l$:
\begin{equation}
\Phi^s_{i}(j)\equiv  \sum_{l\neq i,j}\theta _{i}(l)\left( \sum_{\sigma }\sigma
n_{l\sigma }^{b} \right)   .  \label{phijs}
\end{equation}%
According to Eq. (\ref{sco2}),
each unpaired spinon will contribute to a $2\pi$ vortex to the phase of the superconducting order parameter and their confinement will result in
the superconducting phase coherence.

\begin{widetext}
With the $s$-wave amplitude $\langle\hat{\Delta}_{ij}^a e^{-i\phi^0_{ij}} \rangle$ and phase coherence $\langle F_{ij} \rangle\neq 0$,
the $d$-wave pairing symmetry of the order parameter can be further identified by comparing the phase difference between the two NN bonds,
i.e., $i,i+\hat{x}$ and $i,i+\hat{y}$, as follows. First, one finds
\begin{eqnarray}
F_{ii+\hat{x}}F^*_{ii+\hat{y}} & =& \left(e^{i\sum_{\Delta }A^s} \right)
%e^{i\sum_{l\neq i, i+\hat{x},i+\hat{y}}[\theta _{i+\hat{x}}(l)-\theta _{i+\hat{y}}(l)]
%( \sum_{\sigma }\sigma n_{l\sigma }^{b}) }
\nonumber \\
&\times& e^{i\theta_{i+\hat{x}}(i+\hat{y})\sum_{\sigma }\sigma n_{i+\hat{y}\sigma }^{b}-\theta _{i+\hat{y}}(i+\hat{x})\sum_{\sigma }\sigma
n_{i+\hat{x}\sigma }^{b}}     ,
\label{sco3}
\end{eqnarray}
where, on the right hand side of the first line, the subscript $\Delta$ denotes a summation over the closed path of the links $(i,i +\hat{x})$,
$(i +\hat{x}, i+\hat{y})$, and $(i+\hat{y}, i)$ for the gauge field $A^s$.

Because of the presence of short-range AF order in $|\mathrm{RVB}\rangle $, such that
$\sum_{\sigma }\sigma n_{i+\hat{y}\sigma }^{b}\simeq \sum_{\sigma }\sigma n_{i+\hat{x}\sigma }^{b}$, one has
\begin{equation}
e^{i\theta_{i+\hat{x}}(i+\hat{y})\sum_{\sigma }\sigma n_{i+\hat{y}\sigma }^{b}-\theta _{i+\hat{y}}(i+\hat{x})\sum_{\sigma }\sigma
n_{i+\hat{x}\sigma }^{b}}\simeq e^{i\left[\theta_{i+\hat{x}}(i+\hat{y})-\theta _{i+\hat{y}}(i+\hat{x})\right]\sum_{\sigma }\sigma n_{i+\hat{y}\sigma }^{b}}=-1
\label{d-wave}
\end{equation}
in the second line of Eq. (\ref{sco3}) (by noting $\theta_{i+\hat{x}}(i+\hat{y})-\theta _{i+\hat{y}}(i+\hat{x})=\pm \pi$).
\end{widetext}

Consequently
\begin{eqnarray}
\langle F_{ii+\hat{x}}F^*_{ii+\hat{y}}\rangle \simeq -\left\langle e^{i\sum_{\Delta }A^s} \right \rangle <0 .
\label{sco4}
\end{eqnarray}
[generally the flux produced by $A^s\simeq 0$ within the small loop $\Delta$ in the bracket on the right hand side of Eq. (\ref{d-wave}) is
vanishingly small and the average of such a phase factor will not change the overall sign] and one thus expects a negative sign difference between
$\langle F_{ii+\hat{x}} \rangle$ and $\langle F_{ii+\hat{y}} \rangle$.  Therefore, the electron pairing order parameter is generally $d$-wave like,
which originates from the phase string effect \emph{and} short-range AF correlations as was pointed out previously in Ref. \onlinecite{weng_99}.

\subsubsection{Emergent quasiparticle}

Just like that annihilating a Cooper pair will produce a nonlocal phase factor $F_{ij}$ in Eq. (\ref{sco}), the electron decomposition form (\ref{decomp}) also implies that injecting a bare hole into the system will induce a global (nonlocal) phase shift\cite{Weng_11}. It means that a bare hole state
created by $\hat{c}$ would be different from a low-lying hole excited state by a phase shift $e^{i\hat{\Omega}_{i} }$, i.e,
\begin{eqnarray}
\hat{c}_{i\sigma }|\Psi _{\mathrm{G}}\rangle &=&  e^{i\hat{\Omega}_{i} }\left( \hat{c}_{i\sigma }e^{-i\hat{\Omega}_{i} }\right )|\Psi _{\mathrm{G}}\rangle\nonumber \\
&\rightarrow & e^{i\hat{\Omega}_{i} }\times |\text{low-lying excitation}\rangle.
\end{eqnarray}

In the superconducting phase with $\langle e^{i\hat{\Omega}_i}\rangle \neq 0$, based on the electron fractionalization (\ref{decomp}),
an injected bare hole has an intrinsic fractionalization by
\begin{eqnarray}
\hat{c}_{i\sigma }|\Psi _{\mathrm{G}}\rangle
&\rightarrow & \mathcal{\hat{P}}\left(h_{i}^{\dagger } |\Phi _{h}\rangle \otimes a_{i\bar{\sigma}}^{\dagger
}|\Phi
_{a}\rangle  \otimes e^{i\hat{\Omega}_{i}}|\Phi _{b}\rangle \right)\nonumber \\
&\propto & \langle h_{i}^{\dagger}\rangle \langle e^{i\hat{\Omega}_i}\rangle\left(|\Phi _{h}\rangle \otimes a_{i\bar{\sigma}}^{\dagger
}|\Phi_{a}\rangle  \otimes |\Phi _{b}\rangle \right) .
\end{eqnarray}
Namely, in the presence of holon condensation and phase coherence, a bare hole injected into the superconducting ground state may decay into
an $a$-spinon excitation. In the single-particle spectral function, such a low-lying sharp mode will appear in the antinodal region of $(\pm \pi, 0)$
and $(0,\pm\pi)$ with a spectral weight proportional to the density of holon condensate $|\langle h_{i}^{\dagger}\rangle |^2$ and vanishing above $T_c$
when $\langle e^{i\hat{\Omega}_i}\rangle =0$\cite{Weng_11}.

On the other hand, such a fractionalized quasiparticle mode may not be stable in other momentum region.
As a matter of fact, a quasiparticle excitation as created by the $c$-operator can become a stable mode as a bound state of the
$h^{\dagger}$, $a^{\dagger}$ and the phase shift factor $e^{i\hat{\Omega}}$ near the nodal region, with a BCS-like nodal energy spectral given by
\begin{equation}
E_{\mathbf{k}}=\sqrt{(\epsilon _{\mathbf{k}}-\mu )^{2}+\left( \Delta _{%
\mathbf{k}}\right) ^{2}}.  \label{Ek}
\end{equation}%
Here $\epsilon _{\mathbf{k}}$ is a band spectrum of the original electron with a renormalized hopping integral  $t_{\mathrm{eff}}\propto t(1+\delta )/2$\cite{Weng_11}, $\mu$ is the chemical potential of electrons and a $d$-wave gap function
\begin{equation}
\Delta _{\mathbf{k}}\equiv 2J\sum \limits_{\mathbf{q}}(\cos q_{x}+\cos
q_{y})\Delta _{\mathbf{k+q}}^{SC} . \label{G}
\end{equation}%
Such a quasiparicle mode is presumably coherent in the nodal region at $E_{\bf k}$ which is lower than the gap of an $a$-spinon excitation\cite{Weng_11}. Therefore the present ground state predicts a dichotomy of quasiparticle excitations between the nodal and antinodal regions.

\subsection{Definition of the lower pseudogap phase}
\label{LPPDefinition}

Based on the superconducting ground state  (\ref{scgs-0}), its normal state can be defined by switching off the superconducting coherence in
the wave function. Instead of a conventional Fermi liquid state, new states of matter will emerge in the underdoped regime and  exhibit pseudogap behaviors as to be explored in the following.

According to Eq. (\ref{sco}), the superconducting order parameter
\begin{equation}
\left\langle \hat{\Delta}^{\text{SC}}_{ij}\right\rangle =
\left\langle\hat{\Delta}_{ij}^{0}\right\rangle\left\langle e^{i(1/2)\left[\Phi^s_i(j)+\Phi^s_j(i)\right]}\right\rangle,
\label{scoav}
\end{equation}
requires the simultaneous formation of both pairing amplitude and phase coherence. Hence, the LPP can be defined as a normal state with either
$\langle e^{i(1/2)\left[\Phi^s_i(j)+\Phi^s_j(i)\right]}\rangle=0$ or $\langle \hat{\Delta}_{ij}^{0} \rangle= 0$. In the following we discuss the two
cases separately.

\subsubsection{Lower pseudogap phase I}

The LPP with the pairing amplitude $\langle \hat{\Delta}_{ij}^{0} \rangle\neq 0$ but
\begin{equation}
\left\langle e^{i(1/2)\left[\Phi^s_i(j)+\Phi^s_j(i)\right]}\right \rangle=0
\label{scph-LPP}
\end{equation}
will be defined as the \emph{lower pseudogap phase I} (LPP-I).

An LPP-I state can be naturally obtained as
\begin{equation}
|{\Phi} _{\text{LPP-I}}\rangle = \mathcal{\hat{P}}\left( |\Phi _{h}\rangle \otimes |\Phi
_{a}\rangle  \otimes |\Phi _{b}\rangle\right)_{T>0}   \,.    \label{LPPI}
\end{equation}
Namely, it is a $T>0$ version of the superconducting ground state (\ref{scgs-0}). In other words, the LPP-I state (\ref{LPPI}) is non-superconducting at $T>0$, but reduces to the superconducting ground state at $T=0$. Here $|\Phi _{h}\rangle$ still describes the holon condensation, $|\Phi_{a}\rangle$
describes the $a$-spinons in BCS-like pairing, and $|\Phi _{b}\rangle$ describes the $b$-spinons in bosonic RVB pairing, as to be determined as the mean-field solution below, just like the ground states of Eqs. (\ref{bgs}), (\ref{phia-0}) and (\ref{phirvb}).

It is straightforward to understand why the superconducting phase coherence is thermally disordered in Eq. (\ref{LPPI}) at $T>0$. Note that the
phase $\Phi^s_i$ as defined in Eq. (\ref{phis}), is composed of vortices locking with single $b$-spinons, which are thermally excited in
$|\Phi _{b}\rangle$ once $T>0$, and consequently lead to disordering the phase coherence in Eq. (\ref{scph-LPP}).  Later we shall show that the LPP-I state
actually has a superconducting instability at a {\it finite} $T_c$. At $T<T_c$, the thermally excited spinons in $|\Phi _{b}\rangle$ will further form
loosely bound pairs, due to the confinement force generated by the vortices locking with the $b$-spinons. This will go beyond the mean-field solution (\ref{LPPI}).

\subsubsection{Lower pseudogap phase II}
\label{dfLPPII}

At $T=0$, Eq. (\ref{LPPI}) naturally recovers superconducting phase coherence even at the mean-field level. This is due to the fact that all the $b$-spinons form
\emph{short-range} RVB pairs in the ground state, which ensures the phase coherence in Eq. (\ref{scgs-0}). In order to make phase disordering at $T=0$,
a natural case is that the $b$-spinons form \textit{long-range} RVB pairing such that free neutral spinons can be spontaneously generated without energy cost,
which may happen in the dilute doping boundary with the AFLRO starting to recover\cite{YTQW_11,kou_03}.

But in the following, we shall consider another case of non-superconducting ground state at finite doping, in the {\it absence} of the AFLRO. It corresponds to the case that
the pairing amplitude $\langle \hat{\Delta}_{ij}^{0} \rangle $ vanishes, while the phase coherence
(\ref{sco2}) is still maintained. Here, the pairing of the $a$-spinons is destroyed in $|\Phi_{a}\rangle$ to result in vanishing amplitude of
the Cooper pairing at $T=0$. Such a non-superconducting ground state may be realized by, say, applying strong magnetic fields in the underdoped regime.
It will be called the \emph{lower pseudogap phase II} (LPP-II), which is described by
\begin{equation}
|{\Phi} _{\text{LPP-II}}\rangle = \mathcal{\hat{P}}\left(|\Phi _{h}\rangle \otimes |\Phi
_{a}\rangle_{\Delta^a=0} \otimes |\Phi _{b}\rangle\right) ~ .
\label{LPPII}
\end{equation}

In the ground state of the LPP-II, the holon and $b$-spinon states, $|\Phi _{h}\rangle \otimes |\Phi _{b}\rangle$, will remain essentially the same as
in the superconducting state. But the $a$-spinon pairing disappears in $|\Phi_{a}\rangle_{\Delta^a=0} $. Consequently, a free Fermi gas state of $ |\Phi_{a}\rangle_{\Delta^a=0}$ will dominate the low-energy physics in an LPP-II state, as to be detailed below.

\subsection{Effective Hamiltonian}
\label{EffectiveHamiltonian}

In order to get the effective Hamiltonian for the variational wave function (\ref{scgs-0}), we reexpress the original $t$-$J$ model in Eq. (\ref{HtJ}) in terms of
the fractionalization formalism (\ref{decomp}) as follows
\begin{equation}
H_{t-J}=\mathcal{\hat{P}}(\tilde{H}_{t}+\tilde{H}_{J})\mathcal{\hat{P}}   ,
\label{Eq:Htj}
\end{equation}
where\cite{Weng_11}
\begin{equation}
\tilde{H}_{t}=-t\sum_{\langle ij\rangle\sigma}\left(h^{\dag}_{i}h_{j}e^{iA^{s}_{ij}+ieA_{ij}^e}\right)\left(a^{\dag}_{i\sigma}a_{j\sigma}e^{-i\phi^{0}_{ij}}\right)+\text{h.c.}
\label{Eq:Ht}
\end{equation}
and
\begin{equation}
\tilde{H}_{J}=-\frac{J}{2}\sum_{\langle ij\rangle}\left[(1-n^{h}_{i})(1-n^{h}_{j})(\hat{\Delta}^{s}_{ij})^{\dag}\hat{\Delta}^{s}_{ij}\right]
\label{Eq:HJ}
\end{equation}
in which the bosonic RVB order parameter
\begin{equation}
\hat{\Delta}^s_{ij}\equiv \sum_{\sigma}e^{-i\sigma
A_{ij}^{h}}b_{i\sigma }b_{j\bar{\sigma }} . \label{b-RVBpairing}
\end{equation}

Here, the $h$-holon field formally carries charge $+e$ and couples to the external electromagnetic field $A_{ij}^{e}$ in Eq. (\ref{Eq:Ht}).
The $h$-holons and $b$-spinons are further mutually coupled to each other via the \textrm{U(1)}$\otimes \mathrm{U(1)}$ gauge fields, $A_{ij}^{s}$ and
$A_{ij}^{h},$ respectively, in Eqs. (\ref{Eq:Ht}) and (\ref{Eq:HJ}), which are topological (mutual Chern-Simons) fields as their gauge-invariant flux
strengths in an arbitrary closed (oriented) loop $c$ are constrained to the numbers of spinons and holons within the enclosed area $\Sigma _{c}$:
\begin{equation}
\sum_{c}A_{ij}^{s}=\pi \sum_{l\in \Sigma _{c}}\left( n_{l\uparrow
}^{b}-n_{l\downarrow }^{b}\right) ,  \label{cond1}
\end{equation}%
and
\begin{equation}
\sum_{c}A_{ij}^{h}=\pi \sum_{l\in \Sigma _{c}}n_{l}^{h}.  \label{cond2}
\end{equation}
The origin of such mutual Chern-Simons gauge fields can be traced\cite{weng_97,Weng_11} back to the large gauge (mutual duality) transformation $ e^{i\hat{\Theta}} $ in Eq. (\ref{unitary1}), which precisely incorporates
the nonlocal topological effect of the phase string sign structure hidden in the $t$-$J$ model.

Now we introduce the most essential mean-field order parameter\cite{weng_99,weng_07,Weng_11} for the $t$-$J$ model in the representation of Eqs.  (\ref{Eq:Ht}) and (\ref{Eq:HJ}):
\begin{equation}
\Delta^s\equiv\langle \hat{\Delta}^s_{ij}\rangle
\label{b-RVB}
\end{equation}
with $i=NN(j)$. It is a bosonic RVB order parameter characterizing the spin singlet background, which reduces to the original Schwinger-boson
mean-field order parameter\cite{sfermion} at half-filling, where it well describes quantum AF spin correlations over a wide temperature regime $T_0\sim J/k_B$ ($k_B$ is the Boltzmann coefficient). A finite $\Delta^s$ will persist into the underdoped regime to define a pseudogap phase known as the UPP\cite{GZC2005}, which covers both the superconducting state as well as the
LPP states discussed in this work.

According to Eq. (\ref{decomp}), one expects an additional U(1) gauge symmetry between the $h$-holon and $a$-spinon: $h^{\dagger}_i\rightarrow e^{i\theta_i}h^{\dagger}_i$ and $a^{\dagger}_{i\bar{\sigma}}\rightarrow e^{-i\theta_i}a^{\dagger}_{i\bar{\sigma}}$. The presence of this gauge symmetry implies that a new U(1) gauge field denoted by $A_{ij}^a$  is minimally coupled to $h$-holons and $a$-spinons via $+1$ and $-1$ gauge charge, respectively.
Because of the RVB pairing of the $a$-spinons in the ground state of LPP-I as we will discussed later, this gauge field is generally  `Higgsed'.
In fact, the projection operator $\hat{\mathcal{P}}$ in Eq. (\ref{Eq:Htj}) will result in a general relation\cite{Weng_11}: $(\hat{\Delta}^{a}_{ij})^{\dag}\hat{\Delta}^{a}_{ij}=n^{h}_{i}n^{h}_{j}(\hat{\Delta}^{s}_{ij})^{\dag}\hat{\Delta}^{s}_{ij}$, which ties the RVB pairing of the $b$-spinons with that of the backflow $a$-spinons. If one imposes this constraint by introducing a
Lagrangian multiplier $\gamma$, then a fractionalized effective Hamiltonian can be finally written down as follows
\begin{equation}
\tilde{H}_{\mathrm{eff}}=\tilde{H}_{h}+\tilde{H}_{s}+\tilde{H}_{a}\text{ ,}  \label{heff2}
\end{equation}%
with
\begin{eqnarray}
\tilde{H}_{h}  &=& -t_{h}\sum_{\langle ij\rangle }h_{i}^{\dagger }h_{j}e^{i(A_{ij}^{s}+eA_{ij}^e)}
+\text{h.c.}  \nonumber\\
& & +\lambda_{h}\left(\sum_{i}h^{\dag}_{i}h_{i}-{\delta}N\right)   ,
\label{hh2}
\end{eqnarray}%
\begin{eqnarray}
\tilde{H}_{s} = - J_s\sum_{\langle ij\rangle}\hat{\Delta}^{s}_{ij} + \text{h.c.} +\lambda_{b}
\left(\sum_{i\sigma}b^{\dag}_{i\sigma}b_{i\sigma}-N\right)   ,
\label{hs2}
\end{eqnarray}%
\begin{eqnarray}
\tilde{H}_{a} &=&-t_{a}\sum_{\left \langle ij\right \rangle \sigma }a_{i\sigma }^{\dagger }a_{j\sigma } e^{-i\phi
_{ij}^{0}}+ \text{h.c.} -\gamma\sum_{\langle ij\rangle}\left(\hat{\Delta}
^{a}_{ij}\right)^{\dagger}\hat{\Delta}^{a}_{ij}  \nonumber\\
& & +\lambda_{a}\left(\sum_{i\sigma}a^{\dag}_{i\sigma}a_{i\sigma}-\delta N\right)   ,
\label{ha2}
\end{eqnarray}
where
\begin{eqnarray}
&&J_s= J_{\mathrm{eff}}\Delta^s/2,\\
&&J_{\mathrm{eff}}= J(1-\delta)^2-2\gamma\delta^2,
\end{eqnarray}
and $\lambda_{h/b/a}$ represents the chemical potential for the degrees of freedom of the holon/the
$b$-spinon/the backflow $a$-spinon.

Note that slightly different from Ref. \onlinecite{Weng_11}, here the pairing amplitude of the $a$-spinons is introduced via the Lagrangian multiplier $\gamma$ by implementing the average constraint:  $ (\hat{\Delta}^{a}_{ij})^{\dag}\hat{\Delta}^{a}_{ij}\simeq \delta^2 |\Delta^{s}|^2$. The fluctuations going beyond this mean-field equality can be expressed\cite{Weng_11} by $J \sum_{\langle ij\rangle}\left(\hat{\bf{S}}_i^b\cdot \hat{\bf{S}}^a_j+\hat{\bf{S}}^a_i\cdot\hat{\bf{S}}^b_j\right)$. Such a term is to be omitted in the following since we shall be mainly concerned with the mean-field description of the LPP at lower temperature, where at least one of the degrees of freedom is gapped.

The effective coupling constants, $t_h$ and $t_a$, in $\tilde{H}_{\text{eff}}$ can be determined either variationally or by mean-field approximation, which depend on the bare $t$, $J$
and the doping concentration, as well as the projection operator $\mathcal{\hat{P}}$. In fact, based on the renormalized Gutzwiller approximation scheme\cite{zhang_88}, we have approximately doping-independent $t_a\approx t_h\approx t$. But since the basic sign structure of the $t$-$J$ model has been
rigorously captured by the mutual Chern-Simons gauge fields together with the statistics of the constituent particles, the basic physical behavior that we are concerned with in the long-wavelength, low-energy, should not be qualitatively sensitive to the choices of these effective coupling constants.

Therefore, the hidden ODLRO of $\Delta ^s\neq 0$, without explicitly breaking symmetries, provides the necessary `rigidity' for the present fractionalization to occur. It defines the underdoped regime of the $t$-$J$ model and ensures the validity of the above effective Hamiltonian in the so-called UPP (see below).

\subsection{Fractionalized states as mean-field solutions}
\label{phasediagram}

There are three subsystems in the fractionalized ground state (\ref{scgs-0}), which can be determined at the mean-field level by the effective Hamiltonian $\tilde{H}_{\mathrm{eff}}$ in Eq. (\ref{heff2}). Since the LPP states defined at the beginning of this section are closely related to these mean-field solutions, in the following we discuss each degree of freedom as well as their interplays one by one.

\subsubsection{The holon degree of freedom}

The charge degree of freedom is characterized by the bosonic $h$-holons in $|\Phi_{h}\rangle$ in Eq. (\ref{bgs}). In both
the superconducting and LPP states, the holons as bosons will further experience a Bose-condensation (i.e., $\langle h\rangle\neq0$) according to the definition. Such a holon condensation will provide another hidden rigidity in addition to the two-component RVB pairings of the spinons. As a result, the corresponding low-lying excitation will be `supercurrents' generated from the holon
condensate, which lead to the following unique observable consequences.

Based on $\tilde{H}_h$ in Eq. (\ref{hh2}), the supercurrents contributed by the holon condensate are described by a generalized London equation\cite{WM_02,WQ_06,MW_10}
\begin{equation}
\mathbf{J}_h(\bf{r})=\rho _{s}(\nabla \phi+\mathbf{A}^{s}+e\mathbf{A}^{e}).
\label{eq:le}
\end{equation}
Here, the superfluid stiffness $\rho _{s}\equiv\frac{\rho _{h}}{m_h}$, where $\rho_h$ is the superfluid density of the holons with an effective mass
$m_h=\frac{\hbar^2}{2a^2t_h}$ ($a$ is the lattice constant of the square lattice). Reflecting the Mott physics, $\rho_s \rightarrow 0$ in the
half-filling limit. $\nabla \phi$ ensures the U(1) gauge invariance and satisfies
\begin{equation}
\oint_c d\mathbf{r}\cdot \nabla \phi = 2\pi \times \text {integer}
\label{eq:vor}
\end{equation}%
under the requirement of single-valueness of the holon field.

What is special in Eq. (\ref {eq:le}) is the presence of an emergent `electromagnetic field' vector $\mathbf{ A}^{s}$ in additional to the true
external electromagnetic field $\mathbf{ A}^{e}$. Its gauge-invariant field strength is given by Eq. (\ref{cond1}) in a lattice version, or in the
following continuum version
\begin{equation}
\oint_{c}d\mathbf{r}\cdot \mathbf{A}^{s}(\mathbf{r})=\pi  \int_{\Sigma
_{c}}d^{2}\mathbf{r}\left[ n_{\uparrow }^{b}(\mathbf{r})-n_{\downarrow }^{b}(%
\mathbf{r})\right] ,
\label{eq:fluxAs}
\end{equation}%
where the flux of $\mathbf{A}^{s}$ within an arbitrary loop $c$ on the
left-hand-side is constrained to the enclosed spinon numbers on the right-hand-side, as if a $\pm \pi$ flux-tube is attached to each individual spinon.
Here $n_{\uparrow, \downarrow }^{b}(\mathbf{r})$ denotes the local density of spinons.

Based on Eq. (\ref{eq:le}), each unpaired $b$-spinon will automatically generate a supercurrent vortex, known as a spinon-vortex composite\cite{WM_02,WQ_06}, as follows
\begin{equation}
\oint_{c}d\mathbf{r}\cdot \mathbf{J}_h(\mathbf{r})=\rho_s\left(\pm \pi\right)~  ,
\label{eq:sv}
\end{equation}%
where the loop $c$ encloses a single unpaired spinon (for $\mathbf{ A}^{e}=0$). In other words, besides a conventional minimal $2\pi$-type vortices
given in Eq. (\ref{eq:vor}), the holon condensate can sustain a minimal $\pi$-type vortex, in which a $b$-spinon has to be nucleated at the vortex core.
This `cheap vortex' excitation is one of the most important elementary excitation in the LPP-I.

The supercurrents in Eq. (\ref{eq:le}) will generally cost a kinetic energy according to $\tilde{H}_h$, which is given by\cite{WM_02,WQ_06,MW_10}
\begin{equation}
 L_{h}=\frac{1}{2\rho _{s}}\int d^{2}\mathbf{r} \mathbf{J}_h(\mathbf{r})^{2}.
\label{eq:Lh}
\end{equation}
At low concentration of spinon-vortex excitations, such a `London action' will provide a logarithmic potential between the vortices to make
vortex-antivortex binding and thus superconducting phase coherence. Beyond a critical concentration, the proliferation of these spinon-vortices
will effectively screen out the logarithmic interaction in the same fashion as a Kosterlitz-Thouless (KT) type transition, resulting in the intrinsic LPP-I state at $T>T_c$.

Finally, to ensure the holon condensation, the fluctuations of $\mathbf{A}^{s}$ in Eq. (\ref{eq:le}) should be under control. Indeed,
according to its definition in Eq. (\ref{cond1}), if the underlying neutral $b$-spinons are all in short-ranged RVB-pairing in Eq. (\ref{phirvb}),
$\mathbf{A}^{s}$ can get cancelled out significantly in favor of Bose condensation of the holons as well as the mean-field decoupling between the holon and $b$-spinon
states in the ground state (\ref{scgs-0}). In the following subsection, we shall see that, as a matter of fact, a short-range RVB state in Eq. (\ref{phirvb}) will be in turn caused by the holon condensation self-consistently.

\subsubsection{The $b$-spinon degree of freedom}

The building block of the neutral spin background $|\Phi _{b}\rangle $ is the $b$-spinon, as shown in Eq. (\ref{phirvb}) at $T=0$.
It is governed by $\tilde{H}_s$ in Eq. (\ref{hs2}), in which such neutral spin degrees of freedom are influenced by the doped holes mainly through
the lattice gauge field ${A}^{h}_{ij}$ as well as $J_{\mathrm{eff}}$.  As the holons remain Bose condensed in both the superconducting and LPP states, its gauge-invariant field
strength in Eq. (\ref{cond2}) is basically determined by the local superfluid density $\rho _{h}$ in the following continuum version
\begin{equation}
\oint_{c}d\mathbf{r}\cdot \mathbf{A}^{h}(\mathbf{r}) =\pi \int_{\Sigma _{c}}d^{2}\mathbf{r%
}~\rho _{h}(\mathbf{r}) .
\label{eq:fluxAh}
\end{equation}%
Consequently, ${A}^{h}_{ij}$ becomes a non-dynamic field and $\tilde{H}_s$ in Eq. (\ref{hs2}) can be easily diagonalized, resulting in
a mean-field solution\cite{Weng_11,CW2005} $|\Phi _{b}\rangle $ in Eq. (\ref{phirvb}) [more details are presented in Appendix \ref{mfhs}].

At $T=0$, the mean-field RVB pair amplitude $W_{ij}$ in the state $|\Phi _{b}\rangle $ of Eq. (\ref{phirvb}) has been obtained previously as follows\cite{Weng_11}:
$|W_{ij}|\propto e^{-|\mathbf{r}_{ij}|^2/2\xi^2}$ if $i$ and $j$ belong to opposite sublattices and  $W_{ij}=0$ for two sites on the same sublattice.
Here $\xi=a\sqrt{2/\pi\delta}$ is the corresponding spin-spin correlation length of $|\Phi _{b}\rangle $. Namely, with a finite hole concentration
$\delta$, $|\Phi _{b}\rangle $ describes a short-range AF state with $\xi$ essentially determined by the average hole-hole distance. It is a `ghost' (neutral)
spin liquid state always pinned at half-filling, with the spin excitation gapped at $E_g\propto \delta J$ in the spin-1 excitation spectrum\cite{Weng_11,CW2005}. Note that $\xi$ diverges at $\delta=0$, where $W_{ij}\propto 1/|\mathbf{r}_{ij}|^3$ actually becomes
quasi long-ranged. Correspondingly $|\Phi _{b}\rangle $ exhibits an AF long-range order with $|\mathrm{RVB} \rangle=P_s |\Phi _{b}\rangle$
reproducing\cite{Weng_11} a highly accurate variational ground state energy for the $t$-$J$ model at half-filling\cite{lda_88}.

Thus, the ground state (\ref{scgs-0}) reduces to $|\mathrm{RVB} \rangle$ at half-filling, which is of the same form as the LDA wave function\cite{lda_88} and
naturally restores the AFLRO state of the Heisenberg model. In the superconducting regime, $|\mathrm{RVB} \rangle$ becomes a spin liquid state,
which in turn ensures the phase coherence of Eq. (\ref{sco2}) for the Cooper pairs moving on the `vacuum'
$|\mathrm{RVB} \rangle$.

As discussed above, $|\Phi _{b}\rangle $ will remain the same mean-field solution of $\tilde{H}_s$ in the LPP-I at a finite temperature.
The gapped thermally excited $b$-spinons will then decide the basic thermodynamic properties of the LPP-I as to be detailed in Sec. \ref{PhenomenologyLPP}.

\begin{figure}[h]
  \centering
\includegraphics[width=8cm]{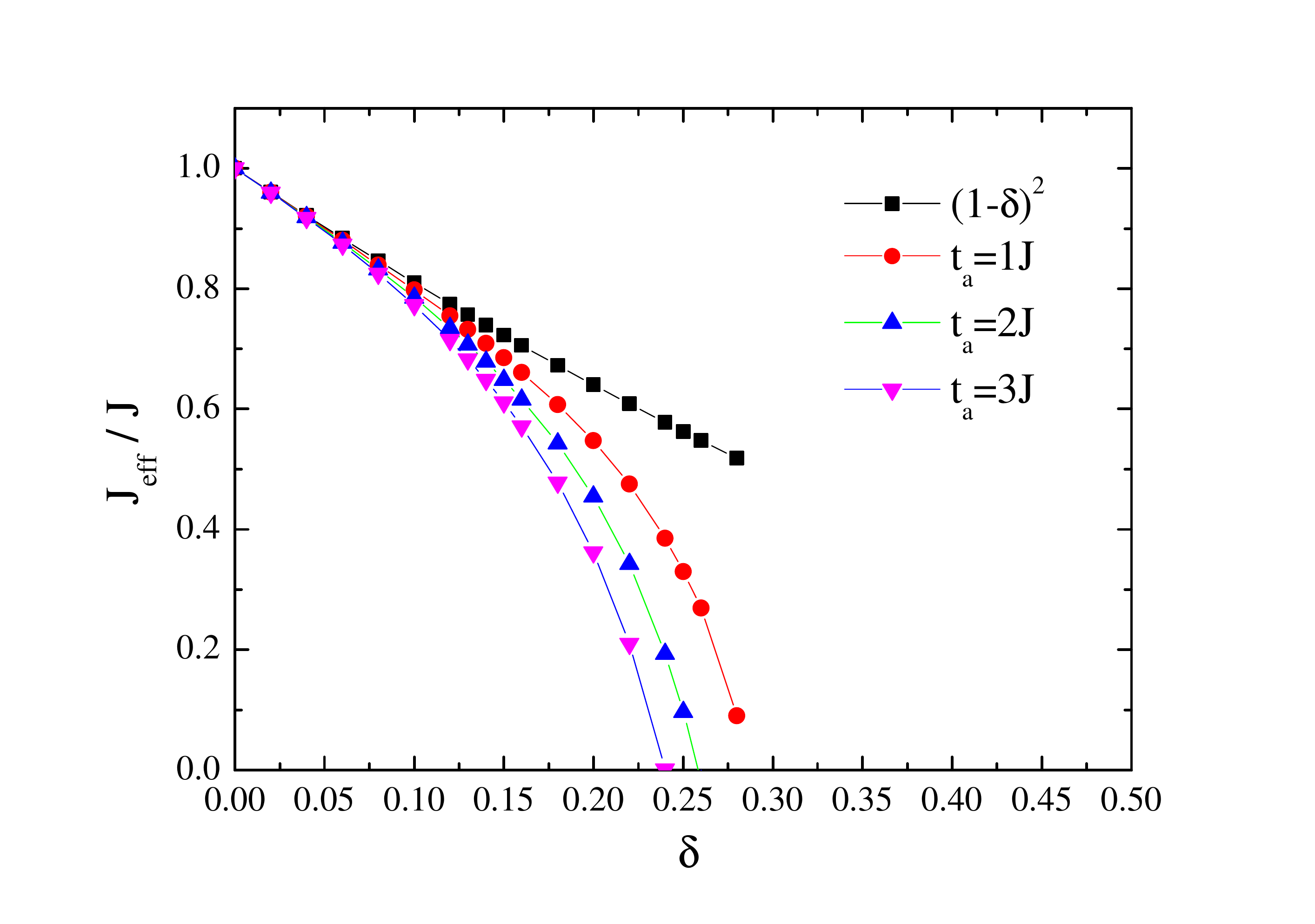}
 \caption{The doping dependence of the renormalized superexchange coupling $J_{\text{eff}}$ at zero temperature, obtained by solving the mean-field self-consistent equations at different values of the parameter $t_a$.  }
  \label{Jeff}
\end{figure}

Besides the gauge field ${A}^{h}_{ij}$, the doping effect will further influence the neutral spinon background by strongly renomalizing the effective superexchange coupling $J_{\mathrm{eff}}$ or $J_s$ in $\tilde{H}_s$ [Eq. (\ref{hs2})]. In a previous approach\cite{GZC2005}, an empirical fitting of $J_{\mathrm{eff}}$ as a function of doping has been used. Here $J_{\mathrm{eff}}$ can be determined self-consistently with the Lagrangian multiplier $\gamma$ after the consideration of the backflow $a$-spinon subsystem (see below). Based on Eqs. (\ref{hs2}) and (\ref{ha2}), the magnitude and doping dependence of $J_{\mathrm{eff}}$ at zero temperature are obtained from the mean-field self-consistent equations as shown in Fig. \ref{Jeff} at different
choices of the parameter $t_a$. Clearly, $J_{\mathrm{eff}}$ decreases monotonically from the bare $J$ with the increase of doping through the interplay between the localized spins (i.e., $b$-spinons) and the itinerant spins (i.e., the backflow $a$-spinons). The latter's density is commensurate with the doping concentration.
Once $J_{\mathrm{eff}}$ is known, the transition temperature $T_0$ for the bosonic RVB order $\Delta^s$ can be determined by
\begin{equation}
T_0=\frac{J_{\mathrm{eff}}}{k_B\ln 3}
\label{T0}
\end{equation}
which defines the UPP boundary\cite{GZC2005,comment1}. By approximately using $J_{\mathrm{eff}}$ calculated at $T=0$ with $t_{a}=2J$ and $J=120\mathrm{meV}$, the crossover temperature $T_0$ (as noted before, $\Delta^s$ does not correspond to a real symmetry breaking) for the UPP is shown in Fig. \ref{phased} (the curve marked by triangles). The critical doping at vanishing  $J_{\mathrm{eff}}$ will separate the underdoped regime from the overdoped regime in the present doped Mott insulator. With $\Delta^s=0$ in the `overdoped' regime, the electron fractionalization discussed so far will no longer be stable at the mean-field level (a Fermi liquid like state may become stabilized at low temperatures as to be discussed later).

\subsubsection{The spinon-vortex composite: An elementary excitation}

Although the superconducting ground state is explicitly fractionalized in terms of three mean-field-type subsystems in Eq. (\ref{scgs-0}), the
aforementioned elementary excitations of two subsystems, i.e., the holon condensate and $b$-spinon RVB background, will be essentially `entangled' by
mutual Chern-Simons gauge fields according to Eqs. (\ref{hh2}) and (\ref{hs2}), resulting in a unique novel excitation: the spinon-vortex composite\cite{WM_02,WQ_06}.
Such a spinon-vortex will be a crucial elementary excitation in characterizing the LPP-I state.

Specifically, the $b$-spinon excitations are created by breaking up RVB pairs in the mean-field spin liquid state $|\Phi _{b}\rangle $.
They are responsible for the pseudogap behavior in the spin degrees of freedom as will be shown in the next section. According to the generalized
London equation (\ref{eq:le}), a superfluid current vortex of vorticity $\pm\pi$ [cf. Eq. (\ref{eq:sv})] will be spontaneously generated around each
$b$-spinon excitation. Then, accompanying the $b$-spinon excitations that are charge-neutral, the associated vortices will play a fundamental role to
result in non-Gaussian-type superconducting fluctuations in the LPP-I state.

\begin{figure}[h]
  \centering
\includegraphics[width=8cm]{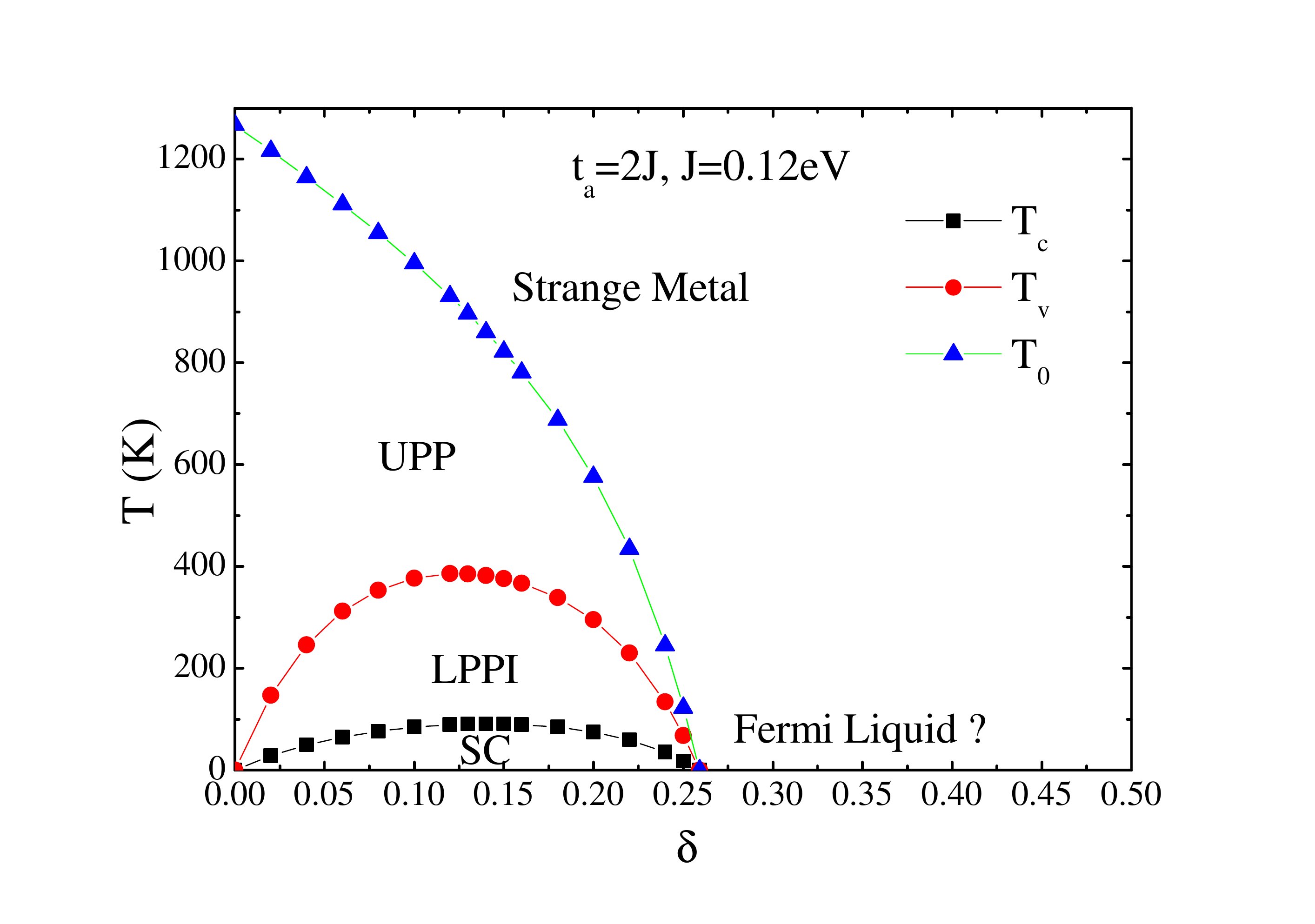}
 \caption{The characteristic temperature scales for the UPP ($T_0$) and the LPP-I ($T_v$) as well as the superconducting (SC) phase ($T_c$) are marked based on the mean field theory of the effective Hamiltonian (\ref{heff2}). Note that in this phase diagram, the AFLRO state at half-filling actually can persist\cite{YTQW_11,kou_03} over a small but finite doping concentration, which will be further investigated elsewhere. The transport and charge dynamics in the so-called strange metal regime at $T>T_0$ have been previously explored in Ref. \onlinecite{GZC2007} based on the effective Hamiltonians (\ref{hh2}) and (\ref{hs2}). In the overdoped regime with $T_0\rightarrow 0$, a possible Fermi-liquid-like instability may occur at low temperature. }
  \label{phased}
\end{figure}

Generally, the spinon-vortex excitations are described by $\tilde{H}_h+\tilde{H}_s$ in Eqs. (\ref{hh2}) and (\ref{hs2}), whose low-energy effective
description is a mutual Chern-Simons gauge theory\cite{YTQW_11,QW_07}, outlined in Appendix \ref{MCSLPP1}. In Sec. \ref{PhenomenologyLPPI} we will
present the basic phenomenology of the LPP-I state governed by such elementary excitations.

In the global phase diagram shown in Fig. \ref{phased}, we present a characteristic temperature $T_{v}$ for the LPP-I, as estimated based on the criterion previously obtained\cite{WQ_06}, i.e., the holon condensation is totally destroyed when the concentration $n_v$ of excited spinon-vortices becomes equal to the concentration $\delta$ of the holons. In the same phase diagram, the superconducting phase transition $T_c$ is determined\cite{MW_10} by Eq. (\ref{eq:tc}), at which the free spinon-vortices form bound pairs  (cf. Sec. IV below), leading to the so-called spinon confinement transition. In Table. \ref{table1}, different phases in the global phase diagram of Fig. \ref{phased} are marked by their corresponding `hidden' ODLROs, where `1' represents a nonzero value of the order parameter and `0' denotes a vanishing value.

\begin{table}
\caption{Hidden ODLROs in the fractionalized degrees of freedom and the corresponding characteristics in the global phase diagram illustrated in Fig. \ref{phased}. Here `1' represents a nonzero value
of the corresponding order parameter and `0' indicates a zero value.}
\begin{tabular}{|c|c|c|c|c|}
  \hline
  Phase & $\Delta^s$  & $\langle h\rangle$  & $\langle F_{ij}\rangle$  &  $\langle \hat{\Delta}^a_{ij}\rangle$   \\  \hline
  SC &   1  &  1   &   1  &  1  \\ \hline
  LPP-I & 1 &  1   &  0   &  1  \\ \hline
  UPP &  1 &   0   &   0  &  1 \\ \hline
  LPP-II & 1  &   1   & 1   & 0  \\  \hline
  Strange Metal &  0  &    0     &     0  &  0       \\  \hline
  Fermi Liquid? &   0   &    0     &     0  &  0       \\
  \hline
\end{tabular}
\label{table1}
\end{table}

\subsubsection{The backflow $a$-spinon}

An emergent fermionic spinon, i.e., the $a$-spinon, is an important component of the ground state (\ref{scgs-0}). It is described by $\tilde{H}_a$ in Eq (\ref{ha2}). The corresponding ground state (\ref{phia-0}) can be obtained as the mean-field solution with $\langle \hat{\Delta}^{a}_{ij}\rangle \neq 0 $, which is similar to a conventional BCS state, but does not carry charge due to its `Meissner' response to the gauge field as pointed out before. Such an itinerant neutral spinon serves as a spin backflow accompanying the hopping of a holon, which describes the hopping effect, in addition to the phase string effect via the mutual Chern-Simons gauge field $A^h$, on the spin degrees of freedom\cite{Weng_11}.

In terms of $\tilde{H}_a$ in Eq (\ref{ha2}), one may write down the corresponding mean-field Hamiltonian as follows
\begin{widetext}
\begin{eqnarray}
\tilde{H}^{MF}_{a} &=&-\left(t_{a}+\gamma\chi^{a}\right)\sum_{\left \langle ij\right \rangle \sigma }e^{-i\phi
_{ij}^{0}}a_{i\sigma }^{\dagger }a_{j\sigma }-\gamma(\Delta^{a})^{*}\sum_{\langle ij\rangle
\sigma }e^{-i\phi
_{ij}^{0}}\sigma a_{i\sigma
}^{\dagger }a_{j\bar{\sigma} }^{\dagger }+{\text{h.c.}} \nonumber\\&&
+\gamma\sum_{\left \langle ij\right \rangle}\left(|\chi^a|^2+|\Delta^a|^2\right)+ \lambda_a\left(\sum_{i\sigma}a_{i\sigma}^{\dag}a_{i\sigma}-{\delta}N\right)  \text{ ,}
\label{hamf}
\end{eqnarray}%
\end{widetext}
where $\chi^a\equiv\left<e^{-i\phi^{0}_{ij}}\sum_{\sigma}a^{\dag}_{i\sigma}a_{j\sigma}\right>=(\chi^a)^{*}$ and
$\Delta^a\equiv\left<e^{-i\phi^{0}_{ij}}\sum_{\sigma}\sigma a^{\dag}_{i\sigma}a^{\dag}_{j\bar{\sigma}}\right>=(\Delta^a)^{*}$. Note that in the presence of a $\pi$ flux depicted by $\phi^{0}_{ij}$, we have found that the $s$-wave solution of $\langle \hat{\Delta}^{a}_{ij}e^{-i\phi_{ij}^{0}}\rangle $ is always more stable than the $d$-wave one at low doping, in contrast to Ref. \onlinecite{Weng_11}.  Here, for convenience, the gauge of $\phi^{0}_{ij}$ will be chosen such that
\begin{align}
e^{-i\phi^{0}_{i,i+\hat{x}}}=(-1)^{i_{y}+1}, \ \
e^{-i\phi^{0}_{i,i+\hat{y}}}=1
\end{align}
%\textbf{The Hamiltonian (\ref{ha}) can be self-consistently solved, and self-consistent equations are:}
%\begin{eqnarray}
%\left<e^{-i\phi^{0}_{ij}}\sum_{\sigma}\sigma a^{\dag}_{i\sigma}a^{\dag}_{j-\sigma}\right>=\Delta^{a},
%\left<e^{-i\phi^{0}_{ij}}\sum_{\sigma}a^{\dag}_{i\sigma}a_{j\sigma}\right>=\chi^{a},
%\left<\sum_{\sigma}a_{i\sigma}^{\dag}a_{i\sigma}\right>=\delta
%\end{eqnarray}%
so there are two sites in a unit cell, and the two sublattices are defined by:
\begin{align}
 i=\{
 \begin{array}{ccc}
      A & \mathrm{if} & i_{y} {\in} \mathrm{ odd}\\
      B & \mathrm{if} & i_{y} {\in} \mathrm{ even}\\
 \end{array}
\end{align}

By the Fourier transformation:
\begin{align}
a_{I\sigma}^{A/B}
=\frac{1}{\sqrt{N/2}}\sum_{\mathbf{k}}
\exp\left(i\mathbf{k}\cdot\mathbf{R}_{I}^{A/B}\right)
a_{\mathbf{k}\sigma}^{A/B}
\end{align}
Here $\mathbf{R}^{A/B}_{I}$ represents the position of $A/B$ site in the unit cell ``$I$''. Then we get
\begin{eqnarray}
\tilde{H}^{MF}_{a} &=&\sum_{\mathbf{k}}\Psi^{\dag}_{\mathbf{k}} \mathcal{M}_{\mathbf{k}}\Psi_{\mathbf{k}}
+\gamma\left(|\chi^a|^2+|\Delta^a|^2\right)\times2N \nonumber\\
&&+{\lambda_a}N(1-{\delta})   ,
\label{HMFa}
\end{eqnarray}
\begin{widetext}
where $\Psi_{\mathbf{k}}\equiv(\Psi^{A}_{\mathbf{k}},\Psi^{B}_{\mathbf{k}})^{T}$,
$\Psi^{A}_{\mathbf{k}}\equiv(a_{\mathbf{k}\uparrow}^{A},a_{-\mathbf{k}\downarrow}^{A\dag})^{T}$,
$\Psi^{B}_{\mathbf{k}}\equiv(a_{\mathbf{k}\uparrow}^{B},a_{-\mathbf{k}\downarrow}^{B\dag})^{T}$,
and the matrix $\mathcal{M}_{\mathbf{k}}$ is:
\begin{align}
\mathcal{M}_{\mathbf{k}}=
\left[
 \begin{array}{cc}
      (-2\tilde{t}_{a}\cos{k_{x}a}+\lambda_a)\sigma_{z}-2\gamma\Delta^{a}\cos{k_{x}a}\sigma_{x} & -2\tilde{t}_{a}\cos{k_{y}a}\sigma_{z}-2\gamma\Delta^{a}\cos{k_{y}a}\sigma_{x}\\
      -2\tilde{t}_{a}\cos{k_{y}a}\sigma_{z}-2\gamma\Delta^{a}\cos{k_{y}a}\sigma_{x} & (2\tilde{t}_{a}\cos{k_{x}a}+\lambda_a)\sigma_{z}+2\gamma\Delta^{a}\cos{k_{x}a}\sigma_{x}\\
 \end{array}
\right]       ,
\end{align}
in which $\tilde{t}_{a}\equiv{t}_{a}+\gamma\chi^{a}$, $\sigma_{x}$ and $\sigma_{z}$ are the Pauli matrices. Then it is straightforward to diagonalize
the mean-field Hamiltonian Eq. (\ref{HMFa}) and obtain the energy dispersions, $\pm\epsilon^{a}_{\mathbf{k}1}$ and $\pm\epsilon^{a}_{\mathbf{k}2}$, by
\end{widetext}
\begin{eqnarray}
\epsilon^{a}_{\mathbf{k}1}=\sqrt{({\xi^{a}_{\mathbf{k}1}})^2+(\Delta^{a}_{\mathbf{k}})^2},
\epsilon^{a}_{\mathbf{k}2}=\sqrt{({\xi^{a}_{\mathbf{k}2}})^2+(\Delta^{a}_{\mathbf{k}})^2},
\end{eqnarray}
where $\xi^{a}_{\mathbf{k}1}=-2\tilde{t}_{a}\sqrt{\cos^2{{k_x}a}+\cos^2{{k_y}a}}+\lambda_a$, $\xi^{a}_{\mathbf{k}2}=
2\tilde{t}_{a}\sqrt{\cos^2{{k_x}a}+\cos^2{{k_y}a}}+\lambda_a$, $\Delta^{a}_{\mathbf{k}}=2\gamma\Delta^{a}\sqrt{\cos^2{{k_x}a}+\cos^2{{k_y}a}}$
due to the $\pi$-flux. The mean-field free energy reads
\begin{eqnarray}
\tilde{F}_{a}^{MF}&=&-\frac{2}{\beta}\sum_{\mathbf{k},\alpha=1}^{\alpha=2}\ln\left[2\cosh\left(\frac{{\beta}\epsilon^{a}_{\mathbf{k}{\alpha}}}{2}\right)\right]
\nonumber\\ &&+\gamma\left(|\chi^a|^2+|\Delta^a|^2\right)\times2N +{\lambda_a}N(1-{\delta})
\end{eqnarray}
where $\beta\equiv\frac{1}{k_{B}T}$. Next, by minimizing this mean-field free energy, i.e.
\begin{eqnarray}
\frac{\partial{\tilde{F}_{a}^{MF}}}{\partial\Delta^{a}}=\frac{\partial{\tilde{F}_{a}^{MF}}}{\partial\chi^{a}}=\frac{\partial{\tilde{F}_{a}^{MF}}}{\partial\lambda_a}=0
\end{eqnarray}
we get the self-consistent equations:
\begin{eqnarray}
\frac{\gamma}{N}\sum_{\mathbf{k},\alpha=1}^{\alpha=2}A_{\mathbf{k}}B_{\mathbf{k}\alpha}=1,\nonumber\\
\frac{1}{N}\sum_{\mathbf{k},\alpha=1}^{\alpha=2}(-1)^{\alpha}\sqrt{A_{\mathbf{k}}}B_{\mathbf{k}\alpha}\xi^{a}_{\mathbf{k}\alpha}=2\chi^{a},\nonumber\\
\frac{1}{N}\sum_{\mathbf{k},\alpha=1}^{\alpha=2}B_{\mathbf{k}\alpha}\xi^{a}_{\mathbf{k}\alpha}=1-\delta \text{ ,}
\end{eqnarray}
where $A_{\mathbf{k}}$ and $B_{\mathbf{k}\alpha}$ are defined as:
\begin{eqnarray}
A_{\mathbf{k}}\equiv\cos^2 k_{x}a +\cos^2 k_{y}a,
B_{\mathbf{k}\alpha}\equiv\frac{\tanh\left(\frac{\beta\epsilon^{a}_{\mathbf{k}{\alpha}}}{2}\right)}{\epsilon^{a}_{\mathbf{k}{\alpha}}}  .
\end{eqnarray}

Hence, the backflow $a$-spinons form a BCS-like pairing state in Eq. (\ref{phia-0}). Due to the $s$-wave nature, in the superconducting state and LPP-I state, they will not contribute to the low-lying dynamics and thermodynamics significantly except for providing a finite hopping integral $t_h$ for the holons and renormalizing $J_{\text{eff}}$. In other words, the $a$-spinons constitute the backborn of the unique fractionalization in Eqs. (\ref{scgs-0}) and (\ref{LPPI}) with a rigidity against the internal U(1) gauge fluctuations. They will also contribute to some unique finite energy dynamics\cite{Weng_11}, which are not the focus of the present work.

Finally, in the LPP-II state defined at the beginning of this section, a special case has been considered, in which the pairing of the $a$-spinons gets suppressed, say, in magnetic vortex cores by strong magnetic fields at low temperature. Here $|\Phi_a\rangle$ in Eq. (\ref{phia-0}) will reduce to a gapless (Fermi liquid) nomal state as a local LPP-II state defined in Eq. (\ref{LPPII}). With vanishing $\Delta^a$ inside the vortex core, the `Meissner' effect or the rigidity due to the $a$-spinon pairing gets destroyed, while the holons still remain Bose condensed in Eq. (\ref{LPPII}). Then the external electromagnetic field $A^e$ will be transferred, via the internal U(1) gauge degree of freedom, from the holon part in Eq. (\ref{hh2}) to solely act on the $a$-spinons. In other words, the $a$-spinons will become \textit{charged} and exposed to the probe of external electromagnetic fields in the LPP-II. On the other hand, with the holon condensation, the internal U(1) gauge fluctuations are still `Higgsed', and thus the $a$-fermions should be quite coherent without feeling strong gauge scattering.

Therefore, a Fermi liquid composed of the $a$-spinons will emerge as a new state of matter in the LPP-II,  which violates the Luttinger theorem for the original electrons without explicitly breaking a global symmetry. Due to the fractionalization in Eq. (\ref{LPPII}), such a new Fermi liquid state is embeded in the backdrop of a superconducting/pseudogap background where the majority of the spin degrees of freedom is still governed by the $b$-spinons. Only in the overdoped regime with $J_\text{eff}\rightarrow 0$, would a different non-superconducting state appear which is beyond the scope of the present work.

\section{Phenomenology of Lower Pseudogap Phases: Experimental Consequences}
\label{PhenomenologyLPP}

In the previous section, we have shown that the low-temperature pseudogap states, i.e., the LPP-I and -II, can be naturally connected to the superconducting ground state (\ref{scgs-0}) as its normal states. In the following, we shall further study the generic spin and charge properties based on the elementary excitations associated with
the fractionalized degrees of freedom in the LPP. On one hand, such anomalous properties can be directly compared to the experimental observations in the cuprates. On the other hand, the unique behaviors of the LPP can reveal the intrinsic non-BCS nature of the superconducting ground state.

\subsection{Lower pseudogap phase I}
\label{PhenomenologyLPPI}

According to the discussion in the previous section, the LPP-I is characterized by three hidden ODLROs, with the superconducting phase coherence destroyed by the thermally
excited spinon-vortices.

In the LPP-I, the spinon-vortex, as a composite of a $b$-spinon binding with a holon supercurrent vortex, plays the essential role in dictating the basic properties. In the following we first focus on the $b$-spinon excitations based on the mean-field description, which determine the spin pseudogap phenomenon.

\subsubsection{Uniform spin susceptibility and specific heat capacity}

\begin{figure}[h]
  \centering
\includegraphics[width=8cm]{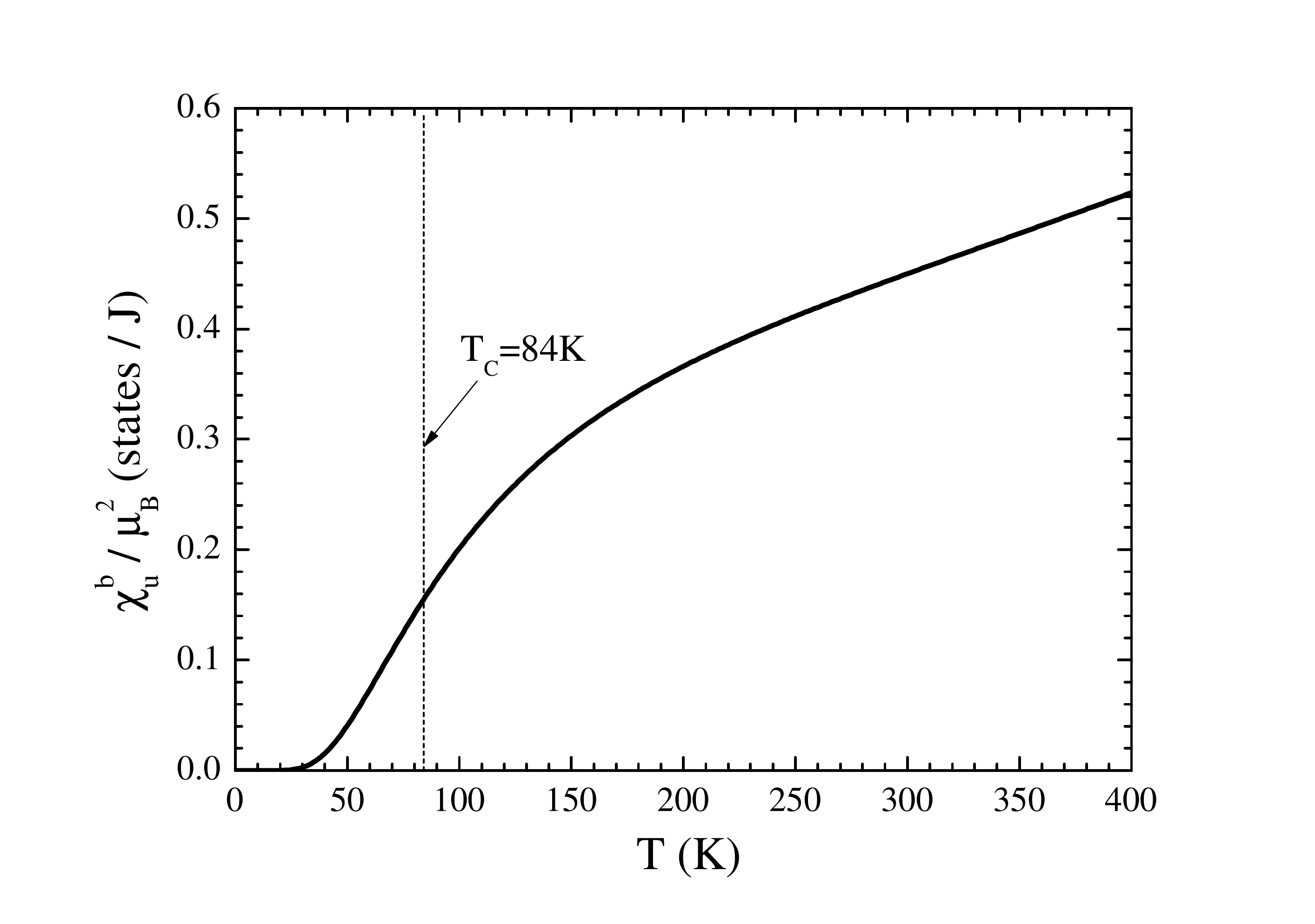}
 \caption{The pseudogap behavior shown by the temperature dependence of the uniform spin susceptibility $\chi_{u}^{b}$ contributed by the $b$-spinons at $\delta=0.1$, obtained with
     $t_a=2J$ and $J=120\text{meV}$. The dashed vertical line marks the characteric temperature $T_c$, below which the LPP-I is no longer stable (see text).}
  \label{chi_s}
\end{figure}

The spin uniform susceptibility $\chi_u^b$ contributed by the $b$-spinons is shown in Fig. \ref{chi_s} at $\delta=0.1$. It is obtained based on
$\tilde{H}_s$ in Eq. (\ref{hs2}) (cf. Appendix \ref{mfhs}):
\begin{eqnarray}
\chi_u^b=\frac{2\mu_{B}^{2}\beta}{N}\sum_{m}n_B(E_{m})\left[n_B(E_{m})+1\right]   ,
\end{eqnarray}
where $E_m$ is the eigen energy of the $b$-spinon excitation,
$\mu_{B}$ is Bohr magneton, and $n_{B}(x)$ is Bose function: $n_{B}(x)\equiv\left[\exp(\beta x)-1\right]^{-1}$.

$\chi_u^b$ exhibits a continuous suppression in magnitude with decreasing temperature over the whole regime of the pseudogap phase ($T<T_0$).
In particular, the vanishing $\chi_u^b$ at low temperature limit is due to a spin gap $E_g$ openned up
in the LPP-I. Note that a dotted vertical line in Fig. \ref{chi_s} marks the superconducting instability of the LPP-I at $T_c$, which is also closely correlated
with $E_g$ (see below).

\begin{figure}
  \centering
\includegraphics[width=8cm]{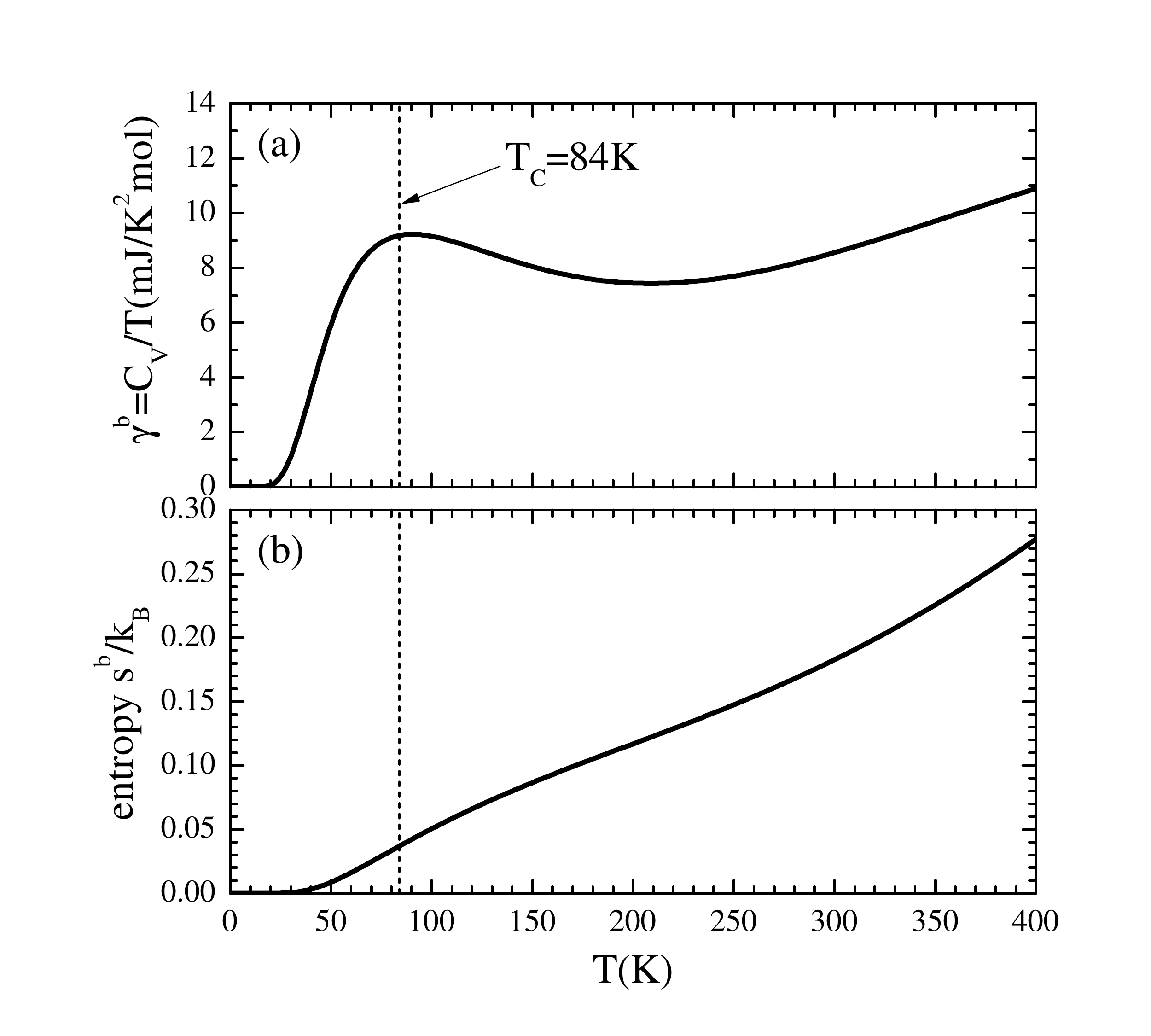}
 \caption{The pseudogap behavior shown by the temperature dependence of (a) the specific heat cofficient $\gamma^{b}$ and (b) the corresponding entropy per site $s^b$ contributed by the $b$-spinon at $\delta=0.1$, with $t_a=2J$ and $J=120\text{meV}$. The dashed vertical line marks the characteric temperature $T_c$, below which the LPP-I is no longer stable (see text). The magnitude of $\gamma^{b}$ is $\thicksim$ 10mJ/K$^2$mol in the LPP-I above $T_c$, which is quite comparable to the experimental data around the optimal doping \cite{J.L. Tallonb2001,Momono}.}
  \label{gamma}
\end{figure}

Similar pseudogap behavior is also exhibited in the spin specific heat capacity of the $b$-spinons, which are shown in Fig. \ref{gamma}(a) at the same doping concentration as in Fig. \ref{chi_s}. The spin specific heat capacity $\gamma^b$ can be expressed by (cf. Appendix B)
\begin{eqnarray}
\gamma^{b}=\frac{1}{N}\sum_{m}\frac{2E_{m}^2}{k_BT^3}n_B(E_m)\left[n_B(E_m)+1\right]   .
\end{eqnarray}

It has been noted that both $\chi^{b}_{u}$ (Fig. \ref{chi_s}) and $\gamma^b$ (Fig. \ref{gamma}(a)) exhibit two distinct `pseudogap' behavior: A slow general decrease with temperature over a wide range down from $T_0$ (defining the UPP as shown in Fig. \ref{phased}) vs. the much steeper suppression at sufficiently low temperatures. The former is due to the formation of the spin RVB pairing (i.e., $\Delta^s\neq 0$), which is already encoded in the mean-field Hamiltonian (\ref{hs2}) and is present even at half-filling, indicating the enhanced AF correlations with reducing $T$.  On the other hand, the latter suppression is due to the fact that a true small spin gap $E_g$ opens up in the LPP-I. It is a direct consequence of the charge condensation in the LPP-I, driven through $A_{ij}^h$ in $\tilde{H}_s$. As discussed before, the LPP-I and UPP are distinguished (cf. Table. \ref{table1}) by that in the latter the bosonic charge carriers (holons) are no longer condensed due to the strong fluctuations of $A_{ij}^s$ in $\tilde{H}_h$, where the generalized London equation (\ref{eq:le}) is not valid anymore. In Figs. \ref{chi_s} and \ref{gamma}, the instability of the LPP-I at $T_c$ is marked by the dotted vertical line, which is also related to $E_g$ by Eq. (\ref{eq:tc}) as to be discussed later. Moreover,  the corresponding entropy per site $s^b$ contributed by the $b$-spinons is shown in Fig. \ref{gamma}(b), which will be further discussed in Sec.\ref{NatureLPP}.

\subsubsection{Longitudinal resistivity}

The longitudinal resistivity $\mathbb{\rho}_e$ in the LPP-I will not be described by a Drude formula, since the quasiparticle excitations are no
longer coherent due to the electron fractionalization\cite{Weng_11}. By contrast, the motion of spinon-vortices will generate a distinct dissipation, which is essentially governed by the dynamics of $b$-spinons\cite{WM_02}.

This is a very unique property in the charge transport described by the mutual Chern-Simons theory\cite{YTQW_11}. In contrast with the Ioffe-Larkin rule in the U(1) gauge theory\cite{LNW_06}, the so-called non-Ioffe-Larkin rule has been previously obtained\cite{YTQW_11}:
\begin{eqnarray}
\mathbb{\rho}_e(\mathbf{q},\omega)=\frac{1}{e^2}\left[\mathbb{\sigma}^{-1}_h(\mathbf{q},\omega)+\pi^2\hbar^2
\mathbb{\sigma}_s(\mathbf{q},\omega)\right]   .
\label{nonIL}
\end{eqnarray}
Here $\sigma_h$ represents the longitudinal holon conductivity, $\sigma_s$ represents the
longitudinal $b$-spinon conductivity with using the SI units: $[\sigma_h]=[\sigma_s]=[\hbar]^{-1}$ (cf. Appendix \ref{nilunits}). At any temperature,
the static conductivity may be obtained by taking the limits, $\mathbf{q}\rightarrow0$ first and then $\omega\rightarrow0$.
Notice that there is no contribution of the backflow $a$-spinon in Eq. (\ref{nonIL}) which is gauge neutral with regard to the mutual Chern-Simons fields and is in a `BCS'  state with regard to the external electromagnetic field.

Due to the condensation of the holon in the LPP-I, we further have $\mathbb{\sigma}^{-1}_h=0$ such that
\begin{eqnarray}
\mathbb{\rho}_e&=&\frac{\pi^2\hbar^2}{e^2}\mathbb{\sigma}_s(\mathbf{q}=0,\omega\rightarrow 0) ,
\label{eq:rhoe}
\end{eqnarray}
%\end{widetext}
where $\mathbb{\sigma}_s(\mathbf{q},\omega)$ denotes the $b$-spinon conductivity. The underlying physical meaning of Eq. (\ref{eq:rhoe}) can be understood by that each spinon excitation behaves like a supercurrent vortex, namely, the spinon-vortex\cite{WM_02,WQ_06}.

At T=0, $\mathbb{\rho}_e\rightarrow 0$ with  $\mathbb{\sigma}_s(\mathbf{q}=0,\omega=0)=0$ as there are no free spinon excitations due to the spin gap
$E_g$. This corresponds to the superconducting ground state. At a finite T, the thermally excited $b$-spinons will make $\mathbb{\sigma}_s$ and thus
$\mathbb{\rho}_e$ in Eq. (\ref{eq:rhoe}) finite, meaning that a non-superconducting phase is naturally realized via the vortex fluctuations associated
with the thermal $b$-spinon excitations. Here the superconducting phase coherence disappears, unless the excited $b$-spinons remain `confined'
within $T<T_c$, paired up via the logarithmic interaction introduced in $L_h$ by Eq. (\ref{eq:Lh})\cite{MW_10}. While a finite $T_c$ will be discussed in the next
subsection, in the following we shall simply assume that such an interaction has been screened such that the LPP-I persists over the whole
low-temperature regime at $T>0$, as described by a finite resistivity in Eq. (\ref{eq:rhoe}).

\begin{figure}[h]
  \centering
\includegraphics[width=8cm]{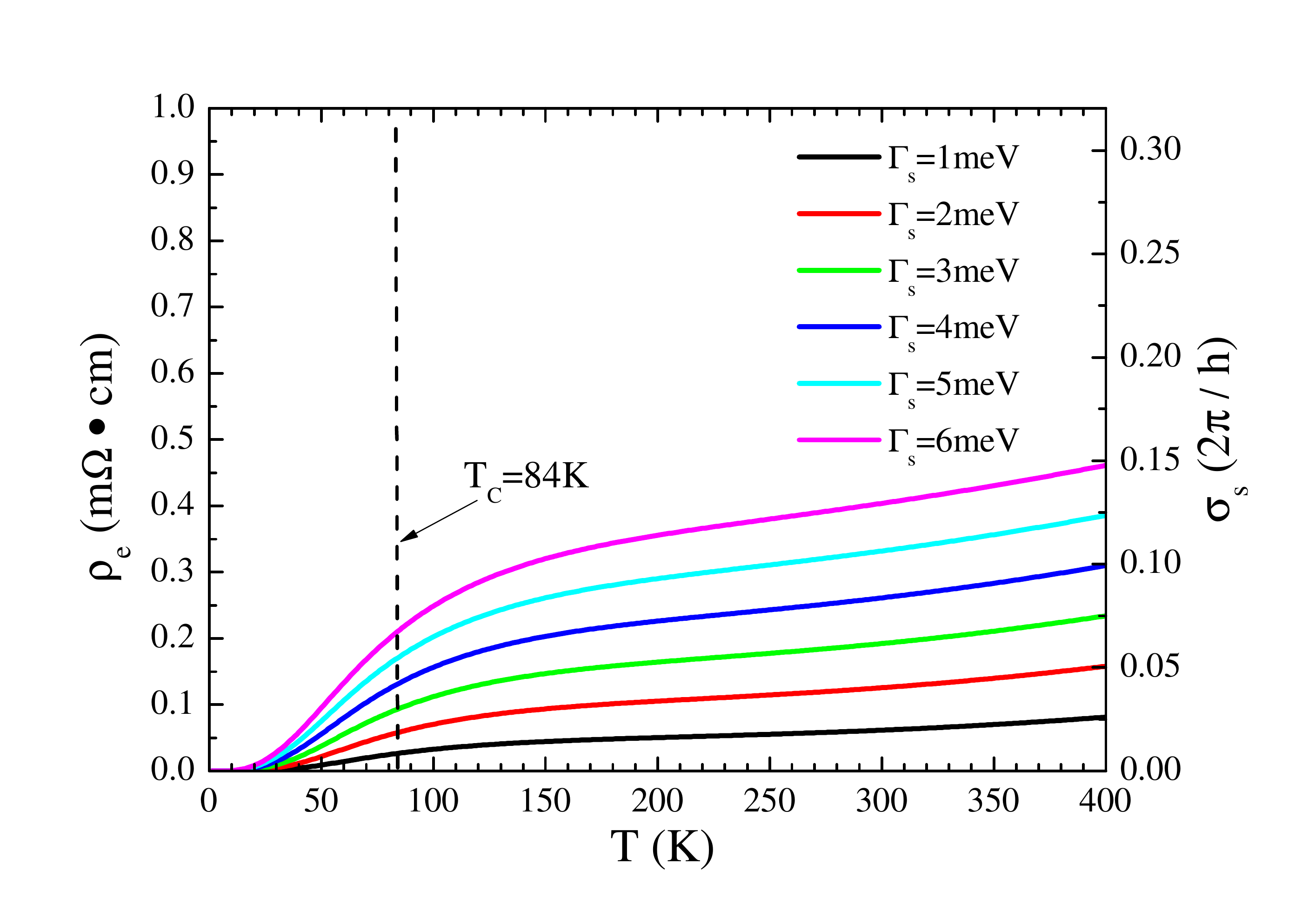}
 \caption{The longitudinal resistivity $\rho_{e}$ in the LPP-I is determined in a non-Drude formula (\ref{eq:rhoe}) by the $b$-spinon conductivity $\sigma_{s}$ contributed by the $b$-spinons. Here $\delta=0.1$, $t_a=2J$ and $J=120\text{meV}$. The parameter $\Gamma_s{\ll}E_g$ specifies the broadening of the spinon spectrum. In order to make comparision with the cuprates,  $\rho_{e}$ is obtained by multiplying the 2D resistance by a lattice constant along the c axis: $d=7.7${\AA}. Here the magnitude of $\rho_{e}$ above $T_c$ is in the range of 0.1m $\Omega\cdot$cm $\thicksim$ 0.5m$\Omega\cdot$cm, which is comparable to the experimental data\cite{W. F. Peck1992}. }
  \label{spinconductivity}
\end{figure}

In the LPP-I, the $b$-spinons are deconfined and described by the mean-field $\tilde{H}_s$. The spinon
conductivity $\sigma_s$ can be calculated by the Kubo formula as given by Eq. (\ref{sigmass}) in Appendix \ref{conductivity}. Figure \ref{spinconductivity} shows the temperature
dependence of $\sigma_s$ and thus $\rho_e$ at $\delta=0.1$. (Here $\rho_{e}$ is obainted by the one-layer resistivity multiplied by the lattice constant along the c axis by $d=7.7${\AA}.) The magnitude of the resistivity is quite comparable to the
experimental data around the optimal doping\cite{W. F. Peck1992,Ando2004}.  To obtain this result, we fix the parameters at $t_a=2J$ and $J=120\text{meV}$.
In addition, a small broadening, $\Gamma_s{\ll}E_g$, is introduced in the spectral function for the mean-field spinon energy level (cf.  Appendix \ref{conductivity}):
\begin{eqnarray}
A(m,\omega)=\frac{1}{\pi}\frac{\Gamma_s}{(\omega-E_m)^2+\Gamma_s^2}.
\end{eqnarray}
In Fig. \ref{spinconductivity}, $\rho_{e}$ ($\sigma_s$) vs. $T$ at different choices of $\Gamma_s$ are shown.

\subsubsection{Nernst effect}

Another peculiar transport phenomenon for the LPP-I in the presence of spinon-vortices is a large Nernst signal\cite{WM_02,WQ_06,QW_07}. Physically,
the spinon-vortices will move along an applied temperature gradient, driven by the entropy associated with the spin-$1/2$ free moments centered at
vortex cores. Because of the motion of supercurrent vortices, transverse electric voltage will be spontaneously established, if those vortices have
a net vorticity polarized by the perpendicular magnetic filed, which is known as the Nernst effect.

The Nernst effect is therefore an important signature of the LPP-I state due to the presence of spontaneous spinon-vortices\cite{WM_02}, which are thermally excited to destroy the superconducting phase coherence.

Based on the generalized London equation (\ref{eq:le}), the Nernst coefficient can be expressed by\cite{WM_02}
\begin{eqnarray}
e_{N}=\alpha_{xy}\rho_{e}   ,
\end{eqnarray}
where
\begin{equation}
\alpha_{xy}=\frac{Bs_{\phi}}{\phi_{0}^{2}n_{v}}   .
\end{equation}
Here $B$ denotes the magnetic field strength, $\phi_{0}\equiv hc/2e$ is the flux quantum. The `transport entropy' $s_{\phi}$ comes from the
the spinon with a free S=1/2 moment locking with a supercurrent vortex, given by\cite{WQ_06}
\begin{eqnarray}
s_{\phi}=k_{B}\{\ln{[2\cosh{(\beta\mu_{B}B)}]}-\beta\mu_{B}B\tanh(\beta\mu_{B}B)\}   .
\end{eqnarray}

\begin{figure}[h]
  \centering
\includegraphics[width=8cm]{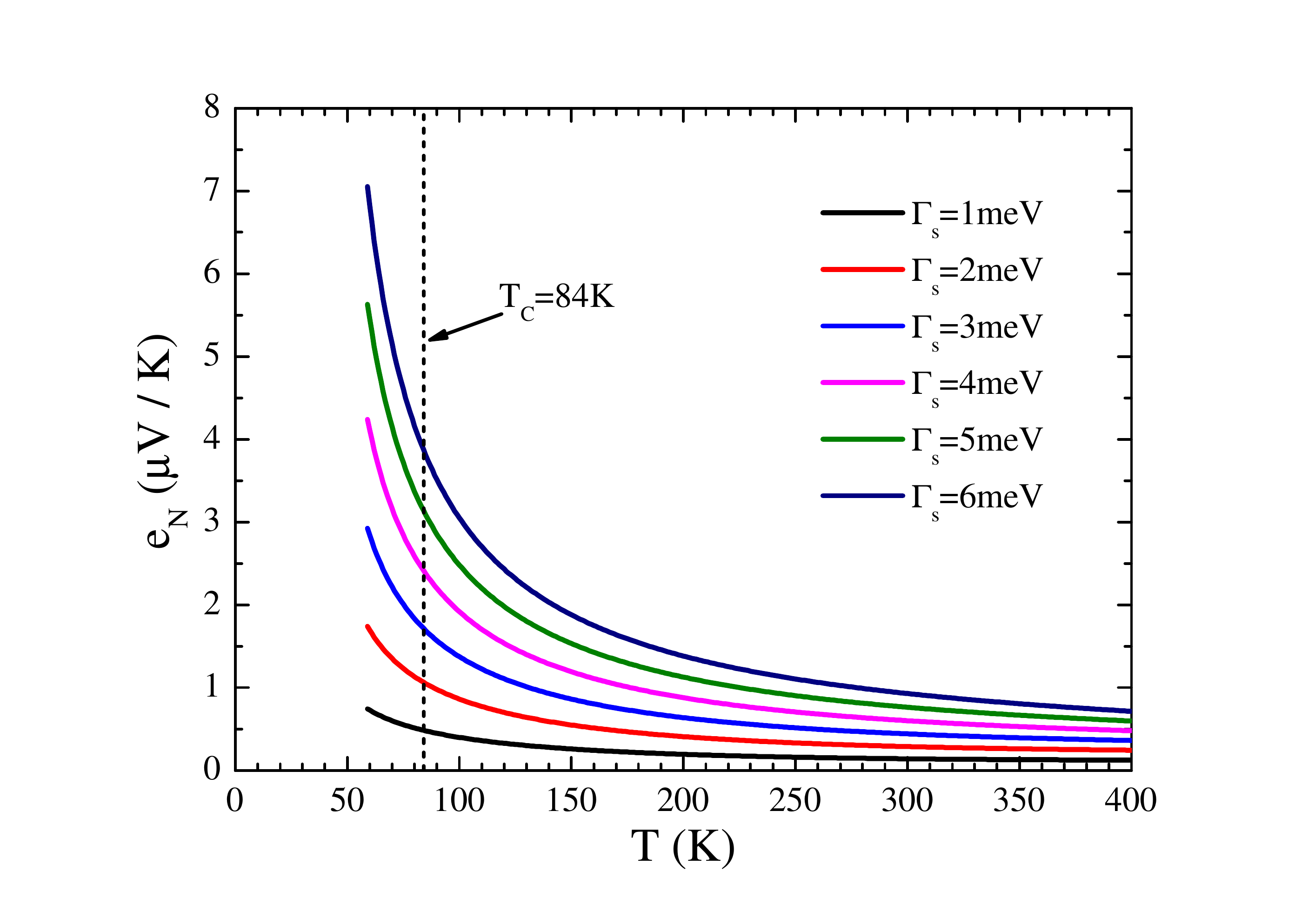}
 \caption{The temperature dependence of the Nernst signal $e_{N}$ at $\delta=0.1$ and $B=20T$, with $t_a=2J$, $J=120\text{meV}$, and different choices of $\Gamma_s$ (cf. Fig. \ref{spinconductivity}). Here the magnitude of $e_{N}$ around $\thicksim$ 1$\mu$V/K above $T_c$  is comparable to the experimental data in the same doping and temperature regime\cite{XZA2000,WYY2001,WYY2006}.}
  \label{eN}
\end{figure}

The temperature dependence of Nernst signal $e_{N}$ at $\delta=0.1$ and $B=20T$ is shown in Fig. \ref{eN}. The magnitude of
the Nernst coefficient is quantitatively comparable to the experimental data\cite{XZA2000,WYY2001,WYY2006}.

\subsubsection{Spin Hall effect}

\begin{figure}
    \centering
    \subfigure[]
    {
        \includegraphics[width=8cm]{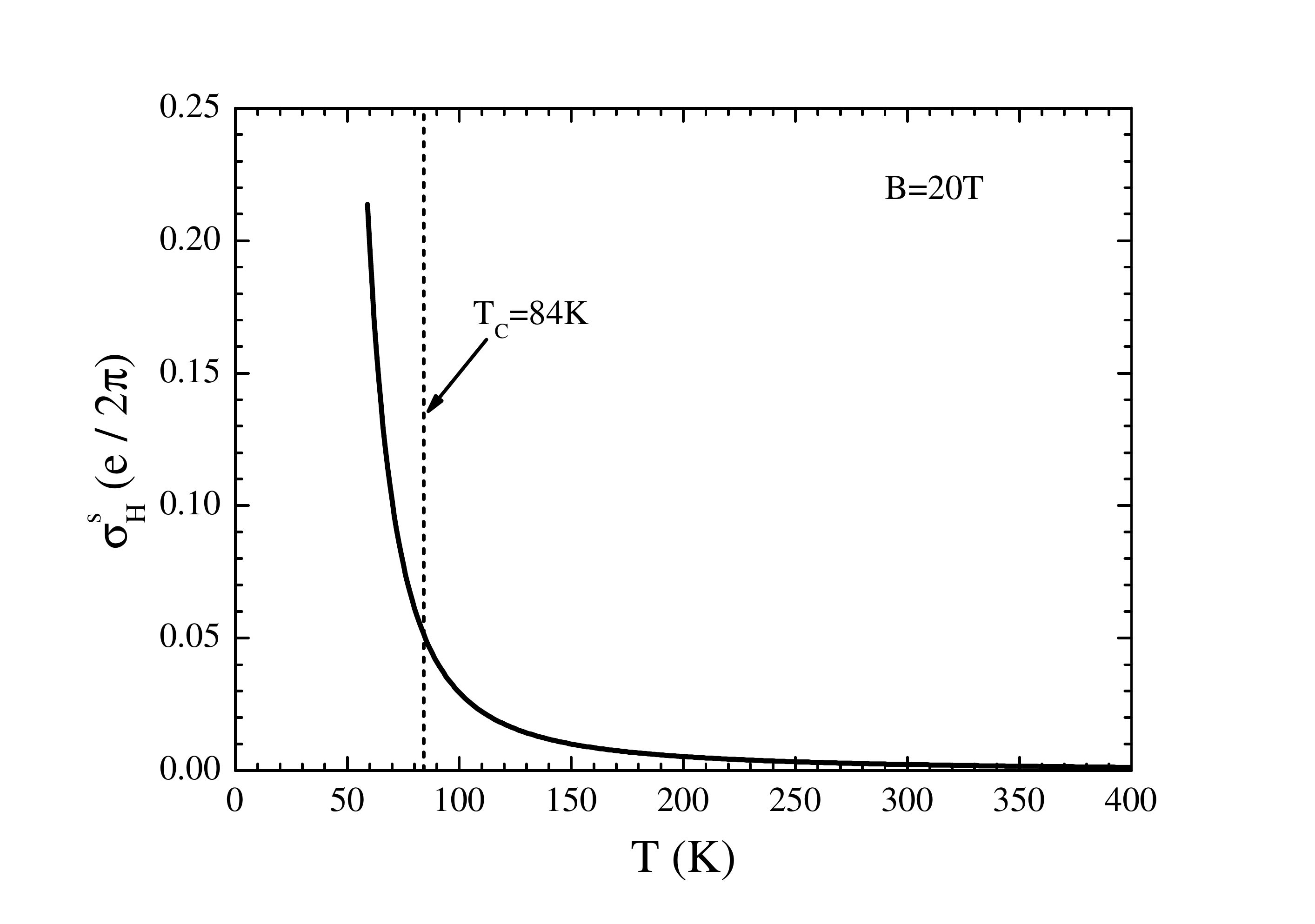}
        \label{spinhallTdp}
    }
    \\
    \subfigure[]
    {
        \includegraphics[width=8cm]{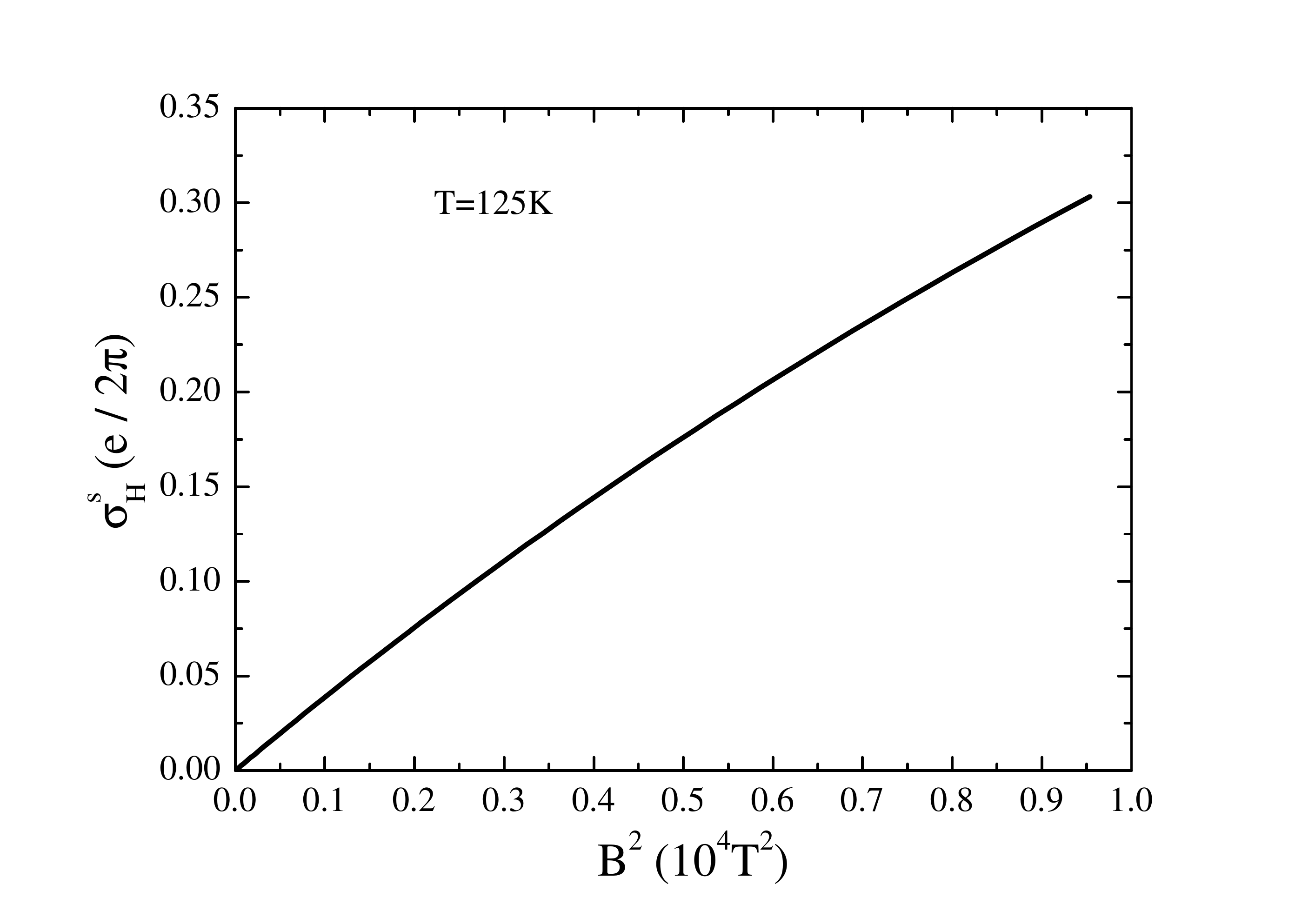}
        \label{spinhallHdp}
    }
    \caption{The temperature (a) and magnetic field (b) dependence of the spin Hall conductivity $\sigma^s_{H}$ at $\delta=0.1$ with
           $t_a=2J$ and $J=120\text{meV}$.}
    \label{spinhall}
\end{figure}

As a unique signature for the presence of spinon-vortices, a dissipationless spin Hall effect has been predicted\cite{KQW_05} for the LPP-I. Physically,
spinon-vortex composites can be driven to move by a perpendicular electric field $E^{e}_{y}$ and consequently
a spin current $J^s_{x}$ is simultaneously generated if the free moments at the centers of the vortex cores are polarized by an external magnetic field $B$ along the $\hat{z}$-axis.
A quantitative prediction is $J^s_{x}=\sigma^s_{H}E^{e}_{y}$, and based on the
generalized London equation (\ref{eq:le}), the spin Hall conductivity can be expressed by\cite{KQW_05}
\begin{eqnarray}
\sigma^s_{H}=\frac{\hbar \chi_u^b}{g\mu_B}\left(\frac{B}{n_{v}\phi_0}\right)^2   ,
\end{eqnarray}
where, the electron $g$-factor $\approx2$. The temperature dependence $\sigma^s_{H}$ at $\delta=0.1$ and $B=20T$ is shown in
Fig. \ref{spinhallTdp} and the magnetic field dependence at $T=125K>T_c$ is shown in Fig. \ref{spinhallHdp}.

\subsubsection{Superconducting instability}

As shown in Figs. \ref{spinconductivity} and \ref{eN}, the resistivity is quickly diminished with a divergent Nernst signal as $T\rightarrow 0$. Such strong non-Gaussian fluctuations in the LPP-I indicate that an intrinsic superconducting instability may happen at a low temperature.

Note that Eq. (\ref {eq:le}) will reduce to a conventional London-like equation describing a superconducting state  if the internal gauge field
$\mathbf{A}^{s}=0$. In fact, the London action (\ref{eq:Lh}) will provide a logarithmic `confinement force' for the spinon-vortices to pair up
at sufficiently low temperatures to make $\mathbf{A}^{s}=0$ at a large length scale.  In other words, the true superconducting condensation is
signalled by the confinement of the $b$-spinons below a critical temperature $T_c$. Correspondingly, the phase coherence condition in
Eq. (\ref{scoav}) is achieved by a vortex-antivortex binding associated with the spinon confinement transition\cite{WM_02,WQ_06,QW_07,MW_10}.
Here one may see the similarity of the current superconducting phase transition to the traditional KT transition for a 2D superfluid system\cite{P. M. Chaikin}.
As a matter of fact, the $T_c$ formula can be similarly obtained as follows\cite{MW_10} (cf. Appendix \ref{mfhs})
\begin{equation}
k_BT_{c}=\frac{E_{g}}{\kappa }
\label{eq:tc}
\end{equation}%
which is controlled by the spin gap $E_g$ with $\kappa\sim6$. Such a $T_c$ has been marked by vertical dotted lines in Figs. \ref{spinconductivity}
and \ref{eN}.

There are several important remarks that concern the nature of the superconducting phase as given below.
First, an ordinary KT transition is driven by conventional $2\pi$ vortices [in the field of $\phi$ of Eq. (\ref {eq:le})].
But here the spinon-vortices involve a vorticity $\pi$.
In contrast to the former, the spinon-vortex-antivortex pairing will not annihilate each other at $T=0$, because of the conserved spinon numbers. Instead,
they form tightly bound vortex-antivortex pairs as the corresponding spinons form RVB pairs in the ground state (\ref{phirvb}). When single spinon-vortices are created by breaking up such RVB pairs, the minimal excitation energy essentially measures the $b$-spinon excitation spectrum (without creating vortices as they are already there in the ground state). Therefore, the spinon-vortex excitation is a `cheap vortex' not only because of the lower vorticity
($\pi$ instead of $2\pi$), but most importantly because of the fact that it still exists as a vortex in the ground state. One can estimate\cite{WQ_06} the lowest energy
to create a pair of spinon-vortices from the ground state as $\simeq E_g\propto \delta J_{\text{eff}}$, which controls $T_c$ as shown in Eq. (\ref{eq:tc}).

Second, corresponding to a finite spin gap $E_g$, the RVB background of the $b$-spinons has a finite spin-spin correlation length
$\xi \sim 1/\sqrt{\delta}$. The contribution of those RVB paired spinons to $\mathbf{A}^{s}$ is thus cancelled out at a length scale larger than $\xi$
or in other words the ground state is a spin liquid state and at the same time a vortex-antivortex binding state. Thermally excited spinons in the LPP-I
are spontaneous vortices which form a vortex liquid\cite{WM_02,WQ_06,QW_07}. Then, when temperature is substantially lower than $E_g/k_B$, only a very small amount of free
spinons get thermally excited. One finds that the logarithmic potential provided by the London action (\ref{eq:Lh}) is sufficient to cause the
confinement of these free spinons and make $\oint_{c}d\mathbf{r}\cdot (\nabla \phi+\mathbf{A}^{s})=0$ at length scales much larger than that of
the spin-spin correlation. Subsequently the superconducting phase coherence is realized. So the precursor of superconductivity in the LPP-I
is closely related to the opening up of the spin gap $E_g$, concomitant with the holon condensation at $T_v$ in Fig. 2.

Third, an important distinction of Eq. (\ref {eq:le}) from the conventional London equation for a BCS superconductor is that a charge $+e$ instead of $2e$ condensate couples
to the electromagnetic field $\mathbf{A}^{e}$ here. Nevertheless, a minimal magnetic flux quantization at $hc/2e$ can be still expected in the present superconducting state\cite{WM_02}. This is because the flux quantization condition is now given by
\begin{equation}
\oint_{c}d\mathbf{r}\cdot {\bf J}_{h} =\oint_{c}d\mathbf{r}\cdot (\nabla \phi+\mathbf{A}^{s}+e\mathbf{A}^{e})= 0,
\label{eq:fluxqu}
\end{equation}
 where, according to Eq. (\ref{eq:fluxAs}), the unit flux quanta of $\oint_{c}d\mathbf{r}\cdot \mathbf{A}^{e}=\pm hc/2e$ (with restoring the full units
of $\hbar$ and $c$) can be still found. The prediction\cite{WM_02} is that each magnetic vortex core must trap a nontrivial zero mode: a free spinon, which leads to
$\oint_{c}d\mathbf{r}\cdot\mathbf{A}^{s}=\pm \pi$.

\subsection{Lower pseudogap phase II: Quantum oscillation}
\label{PhenomenologyLPPII}

In the above, we have discussed a novel magnetic vortex core which traps a free $b$-spinon. At a finite temperature close to $T_c$, spinon-vortices are more easily nucleated by external magnetic fields, right before the thermally excited spinon-vortices destroy the superconducting phase coherence. On the other hand, at sufficiently low-temperature: $T\ll E_g/k_B $, such a novel magnetic vortex may no longer energetically competitive, due to the minimal spin gap $E_g$ in breaking up an RVB pair, as compared to a conventional magnetic vortex of quantization $hc/2e$, which is realized by that the external magnetic field penetrates the $a$-spinon subsystem, thanks to the U(1) gauge freedom associated with decomposing the holon and $a$-spinon as discussed in Sec. \ref{PhenomenologyLPP}. In this case, one finds $\Delta^a\rightarrow 0$ at the vortex magnetic core. Namely, inside the vortex core, one has the gapless $a$-spinon state, i.e., the LPP-II state, instead of trapping a $b$-spinon in the LPP-I. In other words, two types of magnetic vortices are predicted for this non-BCS superconductor, which may appear in different temperatures and doping regimes.

\begin{figure}[h]
  \centering
\includegraphics[width=8cm]{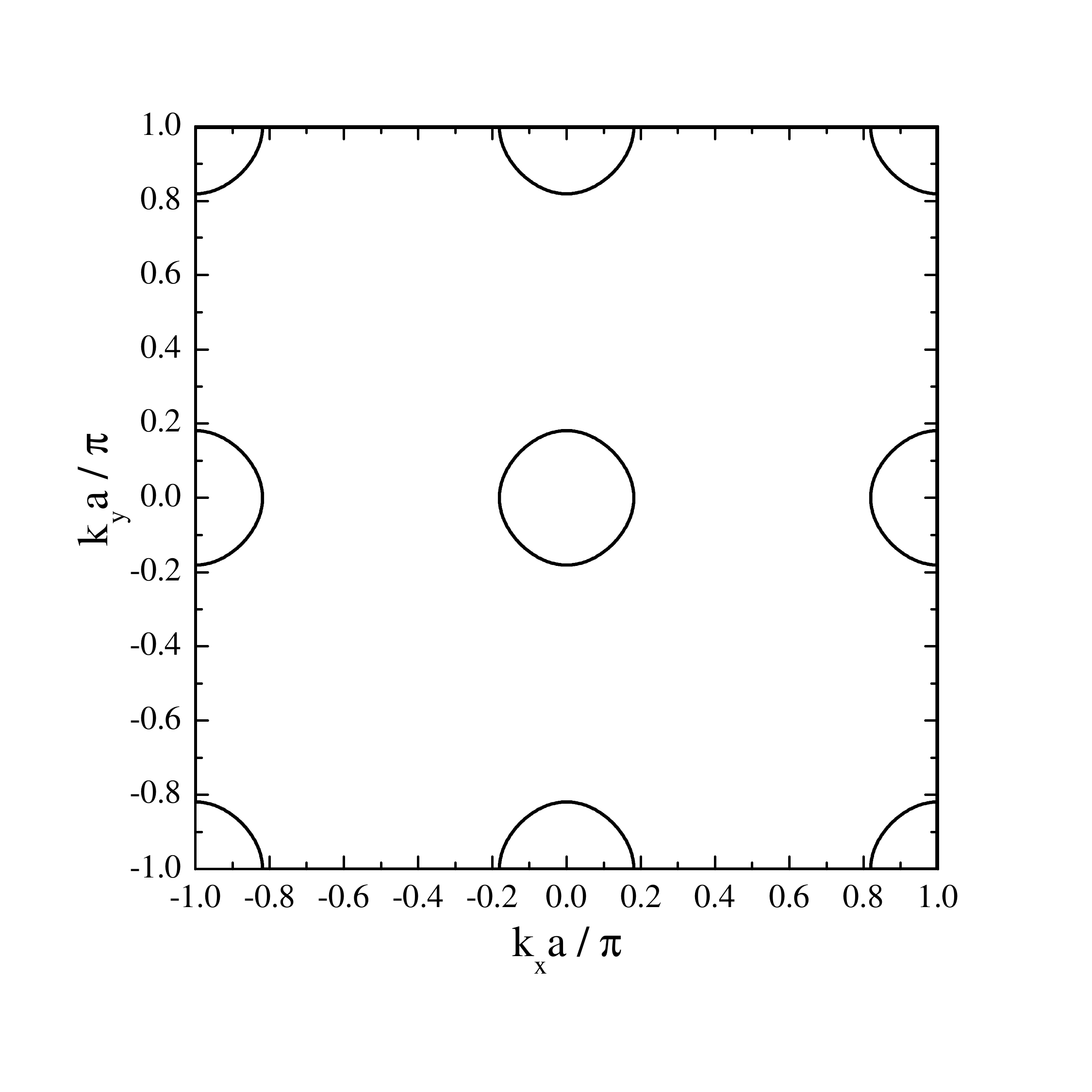}
 \caption{The emergent Fermi surfaces of the $a$-spinon in the LPP-II state, after the superconducting state is suppressed with $\Delta^{a}=0$ by a strong
external magnetic field.  ($\delta=0.1$ and $t_a=2J$.) }
  \label{Eka}
\end{figure}

In the following, instead of justifying its stability, we shall explore the LPP-II state at the mean-field level, which is obtained by turning off $\Delta^{a}$ in $\tilde{H}^{MF}_{a}$ [Eq. (\ref{hamf})]. In the superconducting phase, the $a$-spinons are fully gapped due to its $s$-wave pairing with $\Delta^{a}\neq0$. On the other hand, according to the definition of the
LPP-II in Eq. (\ref{LPPII}),  the $a$-spinons in the LPP-II state will become gapless with $\Delta^{a}=0$. As shown in Fig. \ref{Eka}, the mean-field state discussed in Sec. \ref{phasediagram}4 for the $a$-spinons will reduce to two Fermi pockets around both $(0,0)$ and $(\pi,0)$.

As a matter of fact, in the LPP-II, the DC transport will be solely carried by the $a$-spinons that are in a Landau-Fermi liquid state. To see this clearly, let us first generalize the non-Ioffe-Larkin rule in Eq. (\ref{nonIL}), in which the contribution from the $a$-spinon is not included because the latter remains in the BCS-pairing state in the LPP-I.

By taking account of the internal gauge field $A^a$ that is minimally coupled to $h$-holons and $s$-spinons via $+1$ and $-1$ gauge charge respectively,  we may end up with a general combination rule for transport properties (cf. Appendix \ref{gnonIL}):
\begin{align}
\mathbb{\rho}_e(\mathbf{q},\omega)=\frac{1}{e^2}\left[\mathbb{\sigma}^{-1}_a(\mathbf{q},\omega)+\mathbb{\sigma}^{-1}_h(\mathbf{q},\omega)+\pi^2\hbar^2
\mathbb{\sigma}_s(\mathbf{q},\omega)\right]   \,.
\label{nonILgeneral}
\end{align}
Here $\sigma_a$ is the `conductivity' of $a$-spinons and this general formula is applicable to the whole phase diagram where $J_{\mathrm{eff}}$ is nonzero. In the LPP-I, $\sigma_a=\infty$ (at zero frequency) due to BCS pairings of $a$-spinons, which sends Eq. (\ref{nonILgeneral}) back to Eq. (\ref{nonIL}). Then the aforementioned U(1) gauge field between $a$-spinons and $h$-holons is `Higgsed'. However, in the LPP-II considered here, $a$-spinons form a Fermi liquid since strong magnetic field breaks the BCS-pairing of $a$-spinons inside the vortex core with a finite $\sigma_a$. By noting that $\sigma_h=\infty$ (holon condensation) and $\sigma_s=0$ (RVB pairing of the $b$-spinons) at zero temperature, we have the following formula specific to the LPP-II:
\begin{align}
\mathbb{\rho}_e=\frac{1}{e^2}\mathbb{\sigma}^{-1}_a
\label{nonILgeneralplus}
\end{align}
which indicates that the physical electric transport is merely carried by the $a$-spinons, Namely, the $a$-spinons become charged with vanishing $\Delta^a$ inside the vortex core, which is already discussed in Sec.\ref{phasediagram}4.

At low doping $\delta\ll1$, the energy dispersion around $\mathbf{k}=(0,0)$ at the lower band may be approximatively expressed by
\begin{align}
  \xi^{a}_{\mathbf{k}1}\approx\frac{\hbar^{2}\mathbf{k}^{2}}{2m_{a}}-\lambda_{a}' ,
\end{align}
where the effective mass $m_{a}\equiv\frac{\hbar^{2}}{\sqrt{2}a^{2}\tilde{t}_{a}}$ and the effective
chemical potential is defined by $\lambda_{a}'\equiv2\sqrt{2}\tilde{t}_{a}-\lambda_{a}$.

In 2D, the density of states of the $a$-spinon is given by $N_{a}=\frac{m_{a}}{2\pi\hbar^{2}}$, and the chemical potential can be obtained from the
constraint  $\left<\sum_{I\sigma}a^{A\dag}_{I\sigma}a^{A}_{I\sigma}+\sum_{I\sigma}a^{B\dag}_{I\sigma}a^{B}_{I\sigma}\right>={\delta}N$. The 2D area (denoted by  $\mathscr{A}_{\rm F}$) expanded by each of the Fermi surfaces of the $a$-spinons is
\begin{align}
  \mathscr{A}_{\rm F}=\mathscr{A}_{\rm BZ}\frac{\delta}{4}
\end{align}
where $\mathscr{A}_{\rm BZ}=\frac{4\pi^2}{a^2}$. According to the Onsager relation, we can get the frequency of the quantum oscillation by
\begin{align}
 \mathscr{F}=\frac{{\hbar}c}{2{\pi}e}\mathscr{A}_{\rm F}=\frac{\phi_{0}}{2a^2}\delta.
\end{align}
If $\delta=0.1$, we find the frequency of the quantum oscillation $\mathscr{F}\approx 697$ Tesla, and the magnitude of this result is comparable to the experimental data: $\mathscr{F}^{exp}=(530\pm20) $ Tesla at the similar empiracal doping concentration\cite{QOS2007}.

Apart from the quantum oscillation, the emergent Fermi pockets of the $a$-spinons provide a qualitative explanation for some other experimental consequences in strong magnetic fields at low temperature. For example, a finite value of uniform susceptibility
and a linear-$T$ specific heat capacity at extremely low temperature corresponding to the finite density of state on the Fermi energy have been observed after the superconducting state is suppressed by a strong
external magnetic field\cite{GQZ_10,GQZ_12}. In particular, here the low-$T$ Fermi liquid behavior has been found to be embedded in a larger pseudogap background presumably from the $b$-spinons in the present approach.

\section{Critical comparison with the `plain vanilla' RVB theory and the slave-boson approach}
\label{NatureLPP}

In the previous sections, we have explored the LPP physics based on the superconducting ground state ansatz (\ref{scgs-0}) and shown a systematic agreement with the experiments in the cuprate. In particular, we have emphasized throughout the paper that the pseudog physics has truthfully reflected the non-BCS nature of the superconducting ground state. Since such  superconducting ground state as well as the LPP are obtained based on the $t$-$J$ model, it is very meaningful to make a critical comparison of the present approach with the standard `plain vanilla' RVB theory and the slave-boson approach to the same model.

\subsection{The present superconducting ground state vs. the Gutzwiller projected BCS state}
\label{groundstateansatz}

An alternative ansatz for the superconducting ground state of the $t$-$J$ model is the well-known Gutzwiller projected BCS state proposed by Anderson\cite{pwa_87}. It can be written as
\begin{equation}
|\Psi _{\mathrm{RVB}}\rangle =\hat{P}_{\mathrm{G}}|d\text{-}\mathrm{BCS}%
\rangle  \label{BCS}
\end{equation}%
where $|d$-$\mathrm{BCS}\rangle $ denotes an ordinary $d$-wave BCS state and $%
\hat{P}_{\mathrm{G}}$ is a Gutzwiller projection operator enforcing the
following no double occupancy constraint $\hat{n}_i\leq 1$.
Because of $\hat{P}_{\mathrm{G}}$, the Cooper pairing in $|d$-$\mathrm{BCS}%
\rangle $ reduces to the neutralized RVB pairing\cite{pwa_87} at
half-filling.

Mathematically, in order to implement the no-double-occupancy constraint, there are many choices for a formal fractionalization. For example, one may treat the spinon as fermion and the holon as boson, in the so-called slave-boson decomposition\cite{Barnes1976,Coleman1984,LNW_06} or \emph{vice versa} in the so-called slave-fermion decomposition\cite{sfermion,SWF1989,PALee1989}. In the literature, a popular electron fractionalization is the slave-boson approach\cite{LNW_06}, in which the ground state is obtained with the neutral \emph{fermionic} spinons forming a $d$-wave RVB state $|\Phi _{f}\rangle$ and the holons being in a Bose-condensed state $|\Phi _{h}\rangle$.  Namely,
\begin{equation}
|\Psi _{\mathrm{RVB}}\rangle =C \hat{P}_{hf}\left(|\Phi _{h}\rangle \otimes |\Phi _{f}\rangle \right)~,
\label{sb}
\end{equation}%
where the Gutzwiller projection operator $\hat{P}_{hf}$ implements the no double occupancy constraint on the bosonic holons and fermionic spinons: $n_i^h+n_i^f=1$. In fact,  $\hat{P}_{hf}|\Phi _{f}\rangle\rightarrow  \hat{P}_{\mathrm{G}}|d\text{-}\mathrm{BCS}%
\rangle $ due to the holon condensation, where the RVB and Cooper pairings are not
explicitly distinguished at finite doping. Namely, the `plain vanilla' RVB state is equivalent to the electron fractionalization in the slave-boson formalism.

By contrast, the present superconducting ansatz (\ref{scgs-0}) involves a quite different electron fractionalization from the usual slave-particle decomposition, which is given in Eq. (\ref{decomp}). One may reexpress Eq. (\ref{scgs-0}) in terms of Eq. (\ref{decomp}) as follows
\begin{equation}
|\Psi _{\mathrm{G}}\rangle =e^{i\hat{\Theta}} |{\Phi} _{\mathrm{G}}\rangle ~ ,
\label{unitary1}
\end{equation}
where
\begin{equation}
\hat{\Theta} \equiv -\sum_i n_i^h \hat{\Omega}_{i} ~ ,
\label{unitary2}
\end{equation}
and
\begin{equation}
|{\Phi} _{\mathrm{G}}\rangle \equiv C \exp \left(\sum_{ij}g_{ij}\hat{c}_{i\uparrow }\hat{c}_{j\downarrow } \right) |\mathrm{RVB}
\rangle ~ , \label{frac1}
\end{equation}%
with $g_{ij}=(-1)^{i}\tilde{g}_{ij}$.

Firstly, the bosonic RVB state $|\mathrm{RVB}\rangle$ in Eq. (\ref{frac1}) remains always at half-filling, describing an RVB or neutral spin liquid background defined in Eq. (\ref{lda}). Then doped holes are further introduced by the electron annihilating operators, which are paired in Eq. (\ref{frac1}) with a pairing amplitude $g_{ij}$. Namely, the neutral RVB pairing of spins and the charge BCS-pairing are explicitly separated in Eq. (\ref{frac1}), in contrast to the Gutzwiller projected BCS ground state in Eq. (\ref{BCS}) where the two are not distinguished. Here the no double occupancy constraint is automatically enforced so long as $|\mathrm{RVB}\rangle$ remains singly occupied.

Secondly, the above distinction between the neutral spins and doped holes makes the definition of a nonlocal unitary transformation, i.e., $e^{i\hat{\Theta}}$ in Eq. (\ref{unitary1}), possible. Here with ${n}_i^h$ in Eq. (\ref{unitary2}) as the hole number operator, each doped hole will generate a nonlocal \emph{phase shift} $\hat{\Omega}_{i}$ via Eq. (\ref{unitary2}). It plays a crucial role to regulate the singular sign structure of the $t$-$J$ model at the lattice scale by transforming it into a
large-scale geometric/topological phase shift\cite{weng_97,WWZ_08}. Consequently, in a mean-field-type treatment of $|{\Phi} _{\mathrm{G}}\rangle $, this important sign structure can be accurately retained.

To understand such a sign structure, now imagine a given hole moving through a closed path and then count the geometric (Berry) phase contributed by
$e^{i\hat{\Theta}}$ in Eq. (\ref{unitary1}). Note that, combining with $\Phi^0_{i}$, an $\uparrow$-spin will contribute totally nothing but a
$\downarrow$-spin will give rise to a $2\pi$ phase vortex to $\hat{\Omega}_{i}$ in Eq. (\ref{phif}).  It is then easy to see that all the $\downarrow$ spins
enclosed within the loop will each contribute to a $\pm 2\pi$ phase while $0$ outside the loop. As for those $\downarrow$-spins right on the hole loop,
meaning those $\downarrow$ spins exchanged with the hole during the thinking experiment, will each give rise to a $\pm \pi$ phase, resulting in
a string of signs as a nontrivial geometric phase given by
\begin{equation}
e^{i\hat{\Theta}}\rightarrow (-1)^{N_h^{\downarrow}(c)} \times e^{i\hat{\Theta}}
\label{frac2}
\end{equation}
in which $N_h^{\downarrow}(c)$ denotes the total number of hole-$\downarrow$-spin exchanges on the closed loop $c$. Furthermore, additional
statistical signs can be contributed by the fermionic $\hat{c}$-operators in $|{\Phi} _{\mathrm{G}}\rangle$ of Eq. (\ref{frac1}), i.e.,
\begin{equation}
|{\Phi} _{\mathrm{G}}\rangle\rightarrow (-1)^{N_h^{h}(c)}\times |{\Phi} _{\mathrm{G}}\rangle   ,
\label{frac3}
\end{equation}
where $N_h^{h}(c)$ denotes the total number of exchanges between doped holes in a set of close path $c$ by which holes are exchanged.
Here it is noted that the half-filled spin background $|\mathrm{RVB} \rangle$ will not produce any statistical signs as described by \emph{bosonic}
wavefunction in Eq. (\ref{phirvb}). Then, combining Eqs. (\ref{frac2}) and (\ref{frac3}),  the so-called phase string sign structure\cite{weng_97,WWZ_08} of the $t$-$J$ model is precisely reproduced,
which are both geometric and topological as identified previously for arbitrary doping and temperature on a bipartite lattice of any dimensions\cite{WWZ_08},
and in this procedure, the unitary-transformed representation $|{\Phi} _{\mathrm{G}}\rangle$ becomes `smooth' and locally singular-free to allow
for a mean-field treatment. The unitary transformation in Eq. (\ref{unitary1}) may be also called a mutual-duality transformation.

\subsection{Nature of Mott physics}

It has been well appreciated that the no double occupancy constraint in the $t$-$J$ model is a key ingredient reflecting the basic physics of the doped Mott insulator. In both the `plain vanilla' RVB state in Eqs. (\ref{BCS}) and (\ref{sb}) and the present one in Eqs. (\ref{unitary1}) and (\ref{scgs-0}), such a constraint is implemented.

However, in the present approach, the Mott physics for a doped Mott insulator further means the following:  the fermionic statistical sign structure of the original electrons has been completely changed to the phase string sign structure in the restricted Hilbert space. In short, the Mott physics should be understood as the no double occupancy constraint plus a non-Fermi sign structure.

Consequently, different from the conventional slave-particle scheme, a new electron fractionalization (\ref{decomp}) is found.  It leads to the construction of the new class of ground state (\ref{scgs-0}) or (\ref{unitary1}), which precisely satisfies the sign structure. Moreover, the ground state naturally reduces to the most accurate AF state $ |\mathrm{RVB}\rangle $ (the LDA state\cite{lda_88}) at half-filling, where long-range AF correlations are correctly recovered. It serves as an appropriate starting point to understand the doping problem.

The two superconducting ansatz states in Eqs. (\ref{BCS}) and (\ref{unitary1}), or in the fractionalized form Eqs. (\ref{sb}) and (\ref{scgs-0}), are further distinguished by their dramatically different elemenary excitations.

First of all, they both have bosonic holons in a Bose-condensed state $|\Phi _{h}\rangle$. However, other than this similarity, the rest is so drastically distinct. In particular, the holon condensation in the slave-boson approach as given in Eq. (\ref{sb}) should be destroyed at $T_c$ to result in a pseudogap phase. Correspondingly the charge degree of freedom is characterized by a Bose metal\cite{LNW_06,Fisher} with uncondensed bosonic holons in $|\Phi _{h}\rangle$. By contrat, the holons still remian condensed in the present LPP.

Secondly, in the slave-boson approach, the low-lying elementary excitation (nodal Bogoliubov quasiparticle) is reduced to the $f$-spinon in the superconducting phase based on the fermionic RVB state in $|\Phi _{f}\rangle$ of Eq. (\ref{sb}). By contrast, the Bogoliubov quasiparicle is emergent as a bound state of the holon and $a$-spinon as given in Sec. \ref{d-waveSC}2.

Thirdly, in contrast to the fermionic $f$-spinon in the slave-boson approach, there is a bosonic $b$-spinon in the present state. Such a neutral spin excitation will always induce a supercurrent vortex to form a spinon-vortex composite. In particular, a pair of them form the so-called spin-roton excitation as the unique excitation in the superconducting phase, which determines the superconducting phase transition at a lower temperature than the characteristic temperature of the holon condensation, which decides the LPP-I.

Fourthly, a gapped fermionic $a$-spinon is predicted as a unique feature of the two-component RVB state of the present case in Eq. (\ref{scgs-0}) or (\ref{unitary1}), associated with the spin backflow of the holon hopping. In contrast to the $f$-spinon in the slave-boson approach where its number is equal to the total electron number, here the number of $a$-spinons is commensurate with the holon number.

Finally, we point out that although the LPP state has been expressed in terms of three fractionalized particles, the holon, $b$-spinon, and $a$-spinon, the total entropy contributed by them is not expected to be overcounted because each of the subsystem is in an ODLRO state, in which the entropy is generally suppressed. However, all the hidden ODLROs, including the holon condensation and the spinon pairs, will be melted in a high-temperature (strange metal) regime at $T>T_0$ (cf. Fig. \ref{phased}). It is a natural question if the existence of the three fractionalized particles would lead to an overcounting of the entropy, as is the case for the slave-boson mean-field state in comparison with the high-temperature series expansion results\cite{Putikka1992}. We emphasize that different from a Fermi liquid description of the $f$-spinons in the slave-boson approach, which is the main reason for the overcounting there, in the present case, the $b$-spinons in Eq. (\ref{frac1}) are localized free moments satisfying the Curie-Weisss behavior with the entropy per site bounded by $k_{\text{B}}\ln 2$. Furthermore, the contribution of the $a$-spinons as fermions will be reduced by $\delta/(1-\delta)$ as compared to that of the $f$-spinons in the slave-boson approach. However, a more quantitative comparison to the high-temperature numerical results for the $t$-$J$ model, which is beyond the present scope focusing on the LPP at low temperatures, will be discussed elsewhere.

\subsection{Nature of pseudogap physics}

As noted already, the pseudogap phase in the slave-boson approach is basically a Bose metal, with the fermionic spinons remaining in a $d$-wave RVB state. Namely, the fermionic RVB state $|\Phi _{f}\rangle$ in  Eq. (\ref{sb}) is responsible for the pseudogap properties in the spin degrees of freedom, which should be similar to the superconducting state. On the other hand,  the charge dynamics will be governed by the uncondensed holons of a Bose metal, which is subject to further investigations. Eventually, a strange metal phase is expected at higher temperatures/doping concentrations, after the RVB pairing is destroyed.

In the present approach, the UPP, which has not been discussed in the present paper, corresponds to the above pseudogap phase in the slave-boson approach. However, the LPP, which has been explored in this work, has no correspondence in the slave-boson approach. Specifically, the holons still remain Bose-condensed in the LPP. In fact, all the three subsystems in Eq. (\ref{LPPI}) are still in the ODLROs in the LPP-I as emphasized before.

Hence, the LPP is something unique, as predicted by the superconducting ground state (\ref{scgs-0}) or (\ref{unitary1})-(\ref{frac1}). Here it is distinguished from the superconducting state by the thermally excited unpaired spinon-vortex excitations in the LPP-I or by vanishing RVB pairing of the $a$-spinons in the LPP-II.

The most essential characteristic of the LPP is the opening up of a doping-dependent spin gap $E_g$ as indicated in Figs.\ref{chi_s} and \ref{gamma} at low temperature. Such a spin gap in $|\Phi _{b}\rangle $ of Eq. (\ref{LPPI}) describes a spin liquid with a finite spin correlation length. In particular, $E_g$ vanishes in the dilute hole limit to result in an AFLRO in
$|\mathrm{RVB}\rangle=\mathcal {\hat{P}} |\Phi _{b}\rangle$, which at half-filling becomes a very accurate variational ground state of the Heisenberg Hamiltonian.

A finite $E_g$ is caused by the holon condensation via the mutual Chern-Simons gauge field $\mathbf{A}^h$, which is due to the  altered statistical sign
structure of the $t$-$J$ model explicitly formulated in Eq. (\ref{unitary1}) as the mutual duality transformation. This is absent in the slave-boson approach.

Another important prediction of the mutual duality is that the neutral spin excitations ($b$-spinons) will strongly affect the charge condensate by creating supercurrent vortices, i.e., the spinon-vortices. In the LPP-I, their thermal excitations disorder the superconducting phase coherence, resulting in a large non-Drude resistivity and strong Nernst effect as the
characteristics of non-Gaussian-like superconducting fluctuations as illustrated in Figs. \ref{spinconductivity} and \ref{eN}. Note the reduction of the resistivity in Fig. \ref{spinconductivity} and the divergence of the Nernst siginal in Fig. \ref{eN} as the temperature is lowered below $E_g/k_{\text{B}}$.

Eventually, at a sufficiently low temperature, with the thermally excited spinon-vortices greatly reduced in number due to the spin gap, the confinement of them into vortex-antivortex pairs becomes possible, in a fashion of Kosterlitz-Thouless-type transition, which results in a true superconducting phase coherence below $T\leq T_c$ as controled by the spin gap $E_g$ in Eq. (\ref{eq:tc}).

Finally, even in the zero temperature, a non-superconducting state, i.e., an LPP-II state, can be also realized when the BCS-like pairing of the $a$-spinons in $|\Phi _{a}\rangle $ of Eq. (\ref{phia-0}) is destroyed, say, by strong magnetic fields \emph{before} the occurrence of phase disordering by thermally excited $b$-spinon-vortex excitations (cf. Table. \ref{table1}).
Correspondingly, the Cooper pairing amplitude vanishes to result in a non-superconducting normal state, at least in the magnetic vortex core region.

Note that the $a$-spinon in the superconducting phase and the LPP-I state is charge-neutral as well as gauge-neutral, immune from the mutual Chern-Simons gauge
force between the $b$-spinons and holons. But once in the LPP-II, with vanishing RVB pairing, the $a$-spinons will carry the full charge as the holons are still condensed, and the corresponding Fermi pockets of the fermionic $a$-spinons give rise to quantum oscillation, a Pauli
susceptibility and a linear-$T$ specific heat just like in a typical Fermi liquid.

Therefore, the LPP studied in the present work is a unique low-temperature pseudogap phenomenon, which is not present in the simple slave-boson approach. It is physically related to the rigidity associated with the hidden ODLROs in the fractionalized degrees of freedom in the ground state. Due to the sharp distinctions between the superconducting ground states as well as elementary excitations in the slave-boson and present fractionalization schemes, the nature of the low-temperature pseudogap physics thus differ strongly.

\section{Discussion}\label{sec:conclusion}

In this work, we have intended to understand the pseudogap phenomenon observed in the cuprate superconductor through a model study. Namely, we have explored
the so-called low-temperature pseudogap state in a doped Mott insulator based on the $t$-$J$ model. In addition to its intrinsic superconducting
instability, such a state exhibits a systematic pseudogap behavior in both spin and charge degrees of freedom, as shown by the uniform spin
susceptibility, specific heat, non-Drude resistivity, Nernst effect, as well as the quantum oscillation in strong magnetic fields,
etc. These anomalous properties in the low-temperature pseudogap phase are found to be qualitatively consistent with the experimental measurements in
the cuprates.

As an important lesson that we learned from this study, these pseudogap properties unveil the most essential non-BCS nature of the superconducting
state. Namely, they are hidden in the superconducting ground state as an integral part of it, and start to explicitly manifest once the superconducting
ODLRO is turned off by temperature, magnetic field, or other means. In a conventional BCS state, the superconducting ground state is composed of the
Bloch electrons filling up a Fermi sea and forming the Cooper pairs close to the Fermi energy. The non-BCS superconductivity means that the
normal state is no longer a conventional Fermi liquid dominated by the low-lying Landau quasiparticle excitations. In the present ground state, while the Cooper
pairing of the electrons as the true superconducting ODLRO is still present, the quantum numbers of the electrons are in fact all fractionalized with a
peculiar composite structure. The Mottness, namely, the strong on-site Coulomb repulsion, is the fundamental driving force behind such fractionalization.

We point out that the superconducting state of the doped Mott insulator, either the Gutzwille-projected BCS state (\ref{BCS}) or the present one (\ref{unitary1}), is a natural ground state of pure electrons, without needing an extra `gluon' like phonon in a BCS superconductor. In the latter, a Fermi liquid state is a natural ground state for purely electronic degrees of freedom, which sets in as a `normal state' once the Cooper pairing mediated by phonons is turned off\cite{Gennes}.

In this sense, the proposed superconducting ground states in the doped Mott insulator are the stable infrared fixed point states that essentially control all the anomalous pseudogap
behaviors at finite temperature above $T_c$. In other words, the basic correlations exhibited in high-temperature `normal state' regimes
are already encoded in the ground states, in the specific forms of electron fractionalization,  as shown in Eqs. (\ref{sb}) and (\ref{scgs-0}), respectively.

With both the superconducting ansatz states mentioned above satisfying the no double occupancy constraint of the $t$-$J$ model, the present one has two
advantages: (I) It naturally reduces to the most accurate antiferromagnetic state (the LDA state) at half-filling; (II) It precisely keeps track
of the altered statistic signs (known as the phase string sign structure) of the $t$-$J$ model at finite doping. Then the specific fractionalization
dictated by the new sign structure of the doped Mott insulator leads to a peculiar non-BCS superconducting ground state, which manifects the unique
low-temperature pseudogap behavior once the superconducting coherence is removed. The pseudogap phenomenon here
is thus physically related to the rigidity associated with the hidden ODLROs in the fractionalized degrees of freedom in the ground state, which does not necessarily correspond to any explicit spontaneous symmetry breaking.

Several important issues have not yet been explored in the present work. (A) How the superconducting ODLRO terminates at a finite but sufficiently
low doping? We have pointed out that at half-filling, the antiferromagnetic LDA state is naturally recovered as the ground state. But the
antiferromagnetic order is expected to persist over some very dilute amount of doped holes before the superconducting ground state sets in at zero
temperature. The doped holes have been predicted to be self-localized\cite{localization,YW13} in this non-superconducting regime, and the transitions between the AF and SC phases have been studied in the framework of mutual Chern-Simons gauge theory\cite{YTQW_11}. But more detailed properties like the fate
of the backflow fermionic spinons remain to be investigated and compared with experiment. (B) In the overdoped regime with vanishing $J_{\text{eff}}$, if
the low-temperature pseudogap state will eventually become unstable towards a Fermi liquid state at low temperatures? If the answer is yes, then
how this picture can be reconciled with the non-Fermi sign structure of the $t$-$J$ model\cite{zaanen_09}? If the answer is no, then what would be the non-Fermi-liquid
ground state after the superconductivity disappears beyond a sufficiently large doping in the $t$-$J$ model? In particular, if the Fermi liquid behavior
of the overdoped cuprates corresponds to the so-called Mott collapse\cite{zaanen_09,philips_09} due to a finite Hubbard $U$, it should be already beyond the scope of the $t$-$J$
model. Then what would be the reliable doping regime that the $t$-$J$ model may be relevant to the experiment? (C) In the low-temperature pseudogap
state studied in the present work, the detailed behavior of the quasiparticle excitation remains to be investigated. With the vanishing $d$-wave order
parameter due to the proliferation of the spinon-vortices, the quasiparticles are expected\cite{senthil} to become incoherent with the Fermi arc feature observed
in the ARPES spectral function\cite{shen_03}. But a quantitative study is still absent here. (D) The fractionalized structure should not only be exhibited in the
pseudogap phase, but also be present in the superconducting state, for instance, in the normal core of the magnetic vortex as well as in the excitation
spectra of the bulk. While the fate of the backflow fermionic spinons in a normal core has been studied as in the LPP-II, the contributions of such spinon
excitations, emerging at finite doping, to the dynamic spin susceptibility function and the single-particle spectral function need further
studies.\\

\section*{Acknowledgements} \label{acknowledgement}
We would like to thank V. N. Muthukumar, X.-L. Qi, S.-P. Kou, C.-S. Tian and L. Zhang for the previous collaborations and discussions. This work was supported by the NBRPC grant no. 2010CB923003. P.Y. is supported by the Government of Canada through Industry Canada and by the Province of Ontario through the Ministry of Economic Development \& Innovation.

\begin{widetext}
\appendix

\section{Compact mutual Chern-Simons gauge theory description}
\label{MCSLPP1}

In the main text, the LPP state has been discussed in terms of the effective Hamiltonian (\ref{heff2}) at the mean-field level. To go beyond the mean-field treatment\cite{QW_07}, an effective topological field theory description known as the compact mutual Chern-Simons gauge theory\cite{YTQW_11} will be needed. In the following, such a field-theory description for the low-energy physics of Eqs. (\ref{hh2}) and (\ref{hs2}) is presented.

The holons and $b$-spinons are generally coupled via a pair of mutual Chern-Simons gauge fields in the effective Hamiltonians in Eqs. (\ref{hh2}) and (\ref{hs2}), which represent the most fundamental force originated from the phase string sign structure of the doped Mott insulator as emphasized in the main text. In the following Lagrangian formulation\cite{YTQW_11}, these two degrees of freedom can be expressed generally as
\begin{eqnarray}
\mathcal{L}_h\left[h^{\dag},h;A^{s}_{\mu}\right]=h^{\dag}_{I}(d_{0}-iA_{0}^{s}+\lambda_{h})h_{I}-{t_h}h^{\dag}_{I}
\sum_{\alpha}e^{iA^{s}_{\alpha}}h_{I-\alpha}   ,\label{Lh}
\end{eqnarray}
\begin{eqnarray}
\mathcal{L}_s\left[b^{\dag},b;A^{h}_{\mu}\right]=\sum_{\sigma}b^{\dag}_{i\sigma}(d_0-i{\sigma}A^{h}_{0}+\lambda_{b})b_{i\sigma}-J_s
\sum_{\alpha\sigma}\left(e^{i{\sigma}A^{h}_{i+\hat{\alpha},i}}b^{\dag}_{i+\hat{\alpha}\sigma}b^{\dag}_{i\bar{\sigma}}+\text{h.c.}\right)
,\label{Ls}
\end{eqnarray}
\begin{eqnarray}
\mathcal{L}_{CS}\left[A^{s}_{\mu},\mathscr{N}^{s}_{\mu};A^{h}_{\lambda},\mathscr{N}^{h}_{\lambda}\right]=\frac{i}{\pi}\epsilon^{\mu\nu\lambda}
\left(A^{s}_{\mu}-2\pi\mathscr{N}^s_{\mu}\right)d_{\nu}\left(A^{h}_{\lambda}-2\pi\mathscr{N}^h_{\lambda}\right)\,
,\label{LCS}
\end{eqnarray}
where the bosonic matter field $h/b$ representing holon/spinon field is coupled to the statistic gauge field
$A^{s}_{\mu}/A^{h}_{\mu}$, respectively. The two gauge fields are entangled by the mutual-Chern-Simons term
$\mathcal{L}_{CS}$, where $\mathscr{N}^s_{\mu}/\mathscr{N}^h_{\mu}$ is an
integer field in the compact mutual-Chern-Simons theory. Here $A^{s,h}_{\alpha}$ are the compact link variables with
$A^{s,h}_{\alpha}\in[-\pi,\pi)$ and $A^{s,h}_{0}\in\mathbb{R}$. Notice that $\alpha/\beta$ in the superscript or subscript represents the direction, i.e., $\hat{x}$ or $\hat{y}$ in real space. The total Lagrangian is apparently invariant under the local $U(1){\otimes}U(1)$ gauge transformation up to $\mod{2\pi}$.

The LPP state has been defined by the holon condensation $\langle h\rangle\neq0$. Define
$h_{I}=\sqrt{n_h}e^{i\phi(I)}$ with $n_h=\rho_ha^{2}$. One has
\begin{eqnarray}
\mathcal{L}_h=-iA_{0}^{s}n_{h}+t_{h}n_h(A^{s}_{\alpha})^2
\end{eqnarray}
Here the change of variable: $A^{s}_{\mu}{\rightarrow}A^{s}_{\mu}+d_{\mu}\phi$, has been made and
after this shift, the vector field $A^{s}_{\mu}\in\mathbb{R}$ instead of $[-\pi,\pi)$, which ensures the
correctness of subsequent Gaussian integral.

Summing up the intger field $\mathscr{N}^s_{0}$, one gets the quantization of the gauge field
$\frac{i}{\pi}\epsilon^{0\alpha\beta}d_{\alpha}(A^{h}_{\beta}-2\pi\mathscr{N}^h_{\beta})$ and the following effective Lagrangian
\begin{eqnarray}
\mathcal{L}_{eff}=\mathcal{L}_s+t_{h}n_h(A^{s}_{\alpha})^2+\frac{i}{\pi}\left[\epsilon^{0\alpha\beta}d_{\alpha}(A^{h}_{\beta}-2\pi
\mathscr{N}^h_{\beta})-{\pi}n_{h}\right]A^{s}_{0}+\frac{i}{\pi}\epsilon^{\alpha\mu\nu}\left(A^{s}_{\alpha}-2\pi\mathscr{N}^s_{\alpha}\right)
d_{\mu}\left(A^{h}_{\nu}-2\pi\mathscr{N}^h_{\nu}\right)
\end{eqnarray}

In the LPP, we may separate the spatial components of the gauge field
$A^{h}_{\alpha}$ into two parts:
$A^{h}_{\alpha}=\bar{A}^{h}_{\alpha}+{\delta}A^{h}_{\alpha}$, where
$\bar{A}^{h}_{\alpha}$ depicts the background component which satisfies
\begin{eqnarray}
\epsilon^{0\alpha\beta}d_{\alpha}\bar{A}^{h}_{\beta}={\pi}n_{h}
\end{eqnarray}
and ${\delta}A^{h}_{\alpha}$ represents the fluctuating component. Next we combine the original gauge
field with corresponding integer field:
$\tilde{A}^{h}_{\alpha}={\delta}A^{h}_{\alpha}-2\pi\mathscr{N}^h_{\alpha}$
and $\tilde{A}^{h}_{0}=A^{h}_{0}-2\pi\mathscr{N}^h_{0}$, and thus
the new defined field $\tilde{A}^{h}_{\mu}\in\mathbb{R}$, which ensures
the correctness of subsequent Gaussian integral in the resulting effective Lagrangian
\begin{eqnarray}
\mathcal{L}_{eff}&=&\mathcal{L}_s+t_{h}n_h(A^{s}_{\alpha})^2+\frac{i}{\pi}A^{s}_{0}\epsilon^{0\alpha\beta}d_{\alpha}\tilde{A}^{h}_{\beta}
+\frac{i}{\pi}\epsilon^{\alpha\mu\nu}\left(A^{s}_{\alpha}-2\pi\mathscr{N}^s_{\alpha}\right)d_{\mu}\tilde{A}^{h}_{\nu}
\end{eqnarray}
After intergrating out $A_{\mu}^{s}$, one obtains
\begin{eqnarray}
\mathcal{L}_{eff}&=&\mathcal{L}_s+\frac{1}{4\pi^{2}n_{h}t_{h}}\left(\epsilon^{\alpha\mu\nu}d_{\mu}\tilde{A}^{h}_{\nu}\right)^2
-2i\epsilon^{\alpha\mu\nu}\mathscr{N}^s_{\alpha}d_{\mu}\tilde{A}^{h}_{\nu}
\end{eqnarray}
with a constraint on $\tilde{A}_{\alpha}^{h}$ is $\epsilon^{0\alpha\beta}d_{\alpha}\tilde{A}_{\beta}^{h}=0$:

Here we may ignore the imaginary time-dependence of
$\tilde{A}^{h}_{\alpha}$. Then we arrive at
\begin{eqnarray}
\mathcal{L}_{eff}&=&\frac{1}{4\pi^{2}n_{h}t_{h}}\left(\epsilon^{\alpha\beta0}d_{\beta}\tilde{A}^{h}_{0}\right)^2+\mathcal{L}_s(\tilde{A}^{h}_{0}=0;
\tilde{A}^{h}_{\alpha})-i\tilde{A}^{h}_{0}\left(n_{s}+2\epsilon^{0\alpha\beta}d_{\alpha}\mathscr{N}^s_{\beta}\right),
\end{eqnarray}
where $n_{s}(\mathbf{r}_{i})\equiv\sum_{\sigma}{\sigma}n_{i\sigma}^b=\sum_{\sigma}{\sigma}b_{i\sigma}^{\dag}b_{i\sigma}$.
Define the spinon vorticity
\begin{eqnarray}
q_{sv}(\mathbf{r}_{i}){\equiv}n_{s}(\mathbf{r}_{i})+2\epsilon^{0\alpha\beta}d_{\alpha}\mathscr{N}^s_{\beta}(\mathbf{r}_{i}) .
\end{eqnarray}
After integrating out $\tilde{A}^{h}_{0}$, we finally obtain the effective action which works in both the LPP and SC phases:
\begin{eqnarray}
S_{eff}=S_{s}(\tilde{A}^{h}_{\mu}=0)+S_{sv},
\end{eqnarray}
where $S_{s}(\tilde{A}^{h}_{\mu}=0)=\sum_{x}\mathcal{L}_s(\tilde{A}^{h}_{\mu}=0)$ and the second term is the effective action for the interacting spinon vortices:
\begin{eqnarray}
S_{sv}=\int_{0}^{1/{k_{B}T}}d{\tau}\frac{{\pi}{n_h}{t_h}}{2}\ln\left(\frac{R}{a}\right)\left[\sum_{i}q_{sv}(\mathbf{r}_{i})\right]^2
-\int_{0}^{1/{k_{B}T}}d{\tau}\frac{{\pi}{n_h}{t_h}}{2}\sum_{i{\neq}j}q_{sv}(\mathbf{r}_{i})\ln\left(\frac{|\mathbf{r}_{i}-\mathbf{r}_{j}|}{a}\right)
q_{sv}(\mathbf{r}_{j}).
\label{SV}
\end{eqnarray}

The intrinsic superconducting instability of the LPP-I state at low temperature will be further discussed based on the above mutual Chern-Simons gauge theory formulation in Appendix B below.

\section{Pseudogap behavior and superconducting instability}
\label{mfhs}
The link variables, $A_{ij}^{s}$ and $A_{ij}^{h}$, can be regarded as
mediating the mutual statistics coupling between the charge and spin degrees
of freedom, i.e., the \textquotedblleft mutual semion
statistics\textquotedblright \ entanglement. But in the ground state, these two subsystems can be effectively `disentangled',
with each in a condensed state with an ODLRO of its own. Consequently, the fluctuations around such a `saddle-point' state will become well controlled
and well behaved, just as in all the conventional systems with an ODLRO where the emergent `rigidity' suppresses the violent fluctuations of the
many-body degrees of freedom.

\subsection{Mean-field solution of the $b$-spinon}

The Hamiltonian $\tilde{H}_{s}$ in Eq. (\ref{hs2}) with the mean field approximation can be diagonalized by a
Bogoliubov transformation\cite{weng_99,CW2005}
\begin{eqnarray}
b_{i\sigma}=\sum_{m}(u_{m}\gamma_{m\sigma}-v_{m}\gamma^{\dagger}_{m\bar{\sigma}})w_{m\sigma}(i)
\end{eqnarray}
which results in
\begin{eqnarray}
\tilde{H}_{s}^{MF}=\sum_{m\sigma}E_{m}\gamma^{\dag}_{m\sigma}\gamma_{m\sigma}+J_\text{eff}\left|\Delta^s\right|^2N-2\lambda_bN+\sum_{m}E_{m},
\end{eqnarray}
with
\begin{eqnarray}
u_{m}=\frac{1}{\sqrt{2}}\sqrt{\frac{\lambda}{E_{m}}+1}, \ \  v_{m}=\mathrm{sgn}(\xi_{m})\frac{1}{\sqrt{2}}\sqrt{\frac{\lambda}{E_{m}}-1}
\end{eqnarray}
and
\begin{eqnarray}
E_{m}=\sqrt{\lambda_b^2-\xi_{m}^2},
\end{eqnarray}
where $\xi_{m}$ is the eigenvalue of $w_{m\sigma}(i)$ which is the eigenstate of the equation
\begin{eqnarray}
\xi_{m}w_{m\sigma}(i)=-J_s\sum_{j=\mathrm{NN}(i)}e^{i{\sigma}A^{h}_{ij}}w_{m\sigma}(j),
\end{eqnarray}
with $J_s\equiv{J_\text{eff}\Delta^s}/{2}$. In determining the spinon excitation spectrum $E_{m}$, the self-consistent conditions:
$\langle \hat{\Delta}^s_{ij}\rangle=\Delta^s$ and $\left<\sum_{i\sigma}b^{\dag}_{i\sigma}b_{i\sigma}\right>=N$ have to be used.

Based on the above mean-field solution, the ground state of the $b$-spinons, which is effectively decoupled from the other fractional particles, is given in Eq. (\ref{phirvb}), in which one finds that the RVB pairing amplitude $W_{ij}$ $=0$ if both $i$ and $j$ belong to the same sublattice and
decays exponentially at large spatial separations for opposite sublattice
sites $i$ and $j$\cite{Weng_11}
\begin{equation}
\left \vert W_{ij}\right \vert \propto e^{-\frac{|%
\mathbf{r}_{ij}|^{2}}{2\xi ^{2}}}.
\end{equation}
 Here $\mathbf{r}_{ij}$ is
the spatial distance and $\xi $ is the characteristic pair size determined
by the doping concentration: $\xi =a\sqrt{\frac{2}{\pi \delta }}$. As pointed above, once the $b$-spinons are all
short-range paired up in $|\Phi _{b}\rangle $, the fluctuations of $%
A_{ij}^{s}$ would become negligible and the two subsystems of the
holons and $b$-spinons are decoupled as depicted by $|\Phi _{h}\rangle
\otimes $ $|\Phi _{b}\rangle $ . Note that at half-filling where $\rho _{h}=0$, $\tilde{H}_{s}$ in Eq. (\ref{hs2}) reduces to the
Schwinger-boson mean-field Hamiltonian, which well captures the antiferromagnetic (AF) correlations including the long-range AF order
at $T=0$\cite{lda_88,Weng_11}.

\subsection{Spin degrees of freedom}
Uniform spin susceptibility is defined by
\begin{eqnarray}
\chi_{u}^{b}=\frac{M}{NB}\bigg|_{B\rightarrow0}
\end{eqnarray}
where $B$ is the strength of external magnetic field and the $b$-spinon magnetization is given by
\begin{eqnarray}
M=\mu_{B}\int d\omega
n_{B}(\omega)\sum_{m\sigma}{\sigma}A_{\sigma}(m,\omega),
\end{eqnarray}
with
$A_{\sigma}(m,\omega)=\frac{1}{\pi}\frac{\Gamma_{s}}{(\omega-E_{m\sigma})^2+\Gamma_{s}^2}$
and $E_{m\sigma}=E_{m}-\sigma\mu_{B}B$. One gets
\begin{eqnarray}
\chi_{u}^{b}\approx\frac{2\mu_{B}^{2}\beta}{N}\sum_{m}n_B(E_{m})\left[n_B(E_{m})+1\right].
\end{eqnarray}
at $\Gamma_{s}\ll E_{g}$.

Furthermore, the contribution of the $b$-spinon to the specific heat can be evaluated as follows
\begin{eqnarray}
\gamma^{b}\equiv\frac{C_{V}}{T}=-\frac{1}{N}\frac{\partial^{2}}{{\partial}T^2}\tilde{F}^{MF}_{s},
\end{eqnarray}
where the mean field free energy $\tilde{F}^{MF}_{s}$ is given by
\begin{eqnarray}
\tilde{F}^{MF}_{s}=\frac{2}{\beta}\sum_{m}\ln\left(1-e^{-\beta E_{m}}\right)+J_\text{eff}\left|\Delta^s\right|^2N-2\lambda_bN+\sum_{m}E_{m}.
\end{eqnarray}
Consequently
\begin{eqnarray}
\gamma^{b}=\frac{2}{N}\int
d\omega\frac{\omega^2}{k_{B}T^3}n_B(\omega)\left[n_B(\omega)+1\right]\sum_{m}A(m,\omega)\approx\frac{2}{N}\sum_{m}\frac{E_{m}^2}{k_BT^3}n_B(E_m)\left[n_B(E_m)+1\right].
\end{eqnarray}

\subsection{Superconducting instability}

In the LPP, the holons are always condensed such that the holon conductivity $\sigma_h=0$. Thus, according to the non-Ioffe-Larkin rule in Eq. (\ref{nonIL}), the DC resistivity is essentially determined by the $b$-spinon conductivity $\sigma_s$, which will be evaluated based on the above mean-field solution in Appendix D.

Due to the interaction term in Eq. (\ref{SV}), the residual interaction between the $b$-spinon-vortices can lead to their `confinement' at low temperatures. Namely, at a sufficiently low temperature, the dilute spinon-vortices and spinon-antivortices tend to form bound pairs, which then leads to the true superconducting phase coherence as discussed in the main text.  Indeed, such a spinon confinement will make the $b$-spinon conductivity $\sigma_s$ vanishing such that
\begin{eqnarray}
\rho_{e}&&=\frac{\pi^2\hbar^2}{e^2}\sigma_{s}(\mathbf{q}=0,\omega\rightarrow0)\propto\sigma_{s}(\mathbf{q}=0,\omega\rightarrow0)=0.
\end{eqnarray}

In the following, we briefly discuss such a KT-like vortex-antivortex binding transition based on the mutual Chern-Simons gauge theory outlined in Appendix A.

Note that in the above mean-field solution, an eigen state of the $b$-spinon has a wave-packet wave function like\cite{MW_10}
\begin{eqnarray}
|w_{m\sigma}(\mathbf{r}_{i})|^{2}\simeq\frac{a^{2}}{2{\pi}a^{2}_{c}}\exp{\left(-\frac{|\mathbf{r}_i-\mathbf{R}_m|}{2a^{2}_{c}}\right)}
,\label{wave}
\end{eqnarray}
with a `cyclotron length' $a_{c}\equiv{a}/\sqrt{\pi\delta}$. Here
the degenerate levels are labeled by the coordinates $\mathbf{R}_m$,
the centers of the spinon wave packet, which form a von Neumann
lattice with a lattice constant $\xi_0=\sqrt{2\pi}{a_c}$. So the
effective Lagrangian $S_\text{eff}$ in Appendix A can be further simplified as
\begin{eqnarray}
S_\text{eff}&&{\simeq}S_{s}(\tilde{A}^{h}_{\mu}=0)+S_{sv}\nonumber \\&&\simeq\int_{0}^{1/{k_{B}T}}d{\tau}\frac{E_{g}}{2}\sum_{m}{|q_{sv}({\mathbf{R}_m})|}
-\int_{0}^{1/{k_{B}T}}d{\tau}\frac{{\pi}{n_h}{t_h}}{2}\sum_{\mathbf{R}_{m}{\neq}\mathbf{R}_{m'}}q_{sv}(\mathbf{R}_{m})\ln
\left(\frac{|\mathbf{R}_{m}-\mathbf{R}_{m'}|}{\xi_0}\right)q_{sv}(\mathbf{R}_{m'}),
\end{eqnarray}
where $E_{g}$ denotes the minimal energy gap of the spin-1 excitation. Finally by noting $n_h=\rho_{h}a^2$,
$t_h=\frac{\hbar^2}{2m_{h}a^2}$, and the spin stiffness $\rho_{s}\equiv{\rho_h}/{m_h}$, the effective action is rewritten as
\begin{eqnarray}
S_{eff}=\frac{E_{g}}{2k_{B}T}\sum_{m}{|q_{sv}({\mathbf{R}_m})|}-\frac{{\pi}}{4}\frac{\rho_{s}}{k_{B}T}
\sum_{m{\neq}m'}q_{sv}(\mathbf{R}_{m})\ln\left(\frac{|\mathbf{R}_{m}-\mathbf{R}_{m'}|}{\xi_0}\right)q_{sv}(\mathbf{R}_{m'}).
\end{eqnarray}

Next one can take a standard procedure in dealing
with a conventional KT transition\cite{P. M. Chaikin,MW_10}. Define the reduced
stiffness $K=\rho_s/k_{B}T$ and the effective fugacity of each
spinon vortex
$y\equiv{e}^{-E_g/2k_{B}T}$.

Finally, the differential renormalization group (RG) equations are obtained by\cite{MW_10}
\begin{eqnarray}
\frac{dK^{-1}}{dl}=g^2\pi^3y^2+O(y^4),
\end{eqnarray}
\begin{eqnarray}
\frac{dy}{dl}=(2-\frac{\pi}{4}K)y+O(y^3),
\end{eqnarray}
where, $g=4$ is degeneracy for each site $\mathbf{R}_m$ in the von Neumann lattice due to the time reversal and bipartite lattice symmetries.
It is easy to find that the two RG equations above could also be
obtained if we replace $(K,y)$ by $(K/4,gy)$ in the RG equations of
conventional KT transition. Here $K$ is replaced by $K/4$ because
the unit vorticity of each spinon vortex is $\pi$ instead of $2\pi$
of a conventional vortex; and $y$ is replaced by $gy$ because of the
$g$ degeneracy for each site $\mathbf{R}_m$ in the von Neumann
lattice.

Therefore, the RG flow in the present case is the same as in a conventional KT
transition if we replace $(K,y)$ in the latter by $(K/4,gy)$. The RG equations
result in a fixed point at $K^{*}=8/{\pi}$ and $y^{*}=0$, and there is a separatrix passing through the critical point $K^{-1}=\pi/8,y(l)=0$.
Points above this separatrix flow towards large values of $K^{-1}$ and large values of $y$, in another word, toward the phase with unbound spinon vortices.
Point exactly on the separatrix with $K^{-1}<\pi/8$ flow to the critical point. The starting point of flows is on the line
$y=\exp(-E_{g}/2{k_B}T)=\exp(-E_{g}K/2\rho_s)$. The transition temperature is then determined by the intersection of this line with the separatrix.
The flow for $T<T_c$ is towards the line $y=0$, which means no spinon excitation is allowed below $T_c$, which corresponds to the spinon
confinement in the superconducting phase.
\begin{eqnarray}
\langle q_{sv}(\mathbf{R}_m)q_{sv}(\mathbf{R}_{m'})\rangle&&=-2g^{2}y^{2}\left[\frac{|\mathbf{R}_{m}-\mathbf{R}_{m'}|}{\xi_0}\right]^{-{\pi}K/2}\rightarrow0,
\end{eqnarray}
where, the fugacity $y$ is renormalized to zero when $T<T_c$. The transition temperature $T_c$ determined\cite{MW_10} is given in Eq. (\ref{eq:tc}) in the main text.

\section{Definitions and the units of spinon and holon conductivities}
\label{nilunits}

In the compact mutual Chern-Simons theory, the $b$-spinon/holon/$a$-spinon conductivity\cite{YTQW_11} is defined by
\begin{eqnarray}
\mathbf{j}_s=\sigma_s\mathbf{E}_h, \label{uniteq1}
\end{eqnarray}
\begin{eqnarray}
\mathbf{j}_h=\sigma_h(\mathbf{E}_s+e\mathbf{E}_e+\mathbf{E}_a), \label{uniteq2}
\end{eqnarray}
\begin{eqnarray}
\mathbf{j}_a=-\sigma_a\mathbf{E}_a. \label{uniteq3}
\end{eqnarray}
Here $\sigma_{s/h/a}$ represents the $b$-spinon/$h$-holon/$a$-spinon conductivity, and the vector field $\mathbf{j}_{s/h/a}$ represents the corresponding current. Eq.(\ref{uniteq1}) is due to $b$-spinon, which is coupled to $\mathbf{A}^h$ in Eqs.(\ref{b-RVBpairing}) and (\ref{hs2}). Eq.(\ref{uniteq2}) is due to $h$-holon, which is charged and also coupled to $\mathbf{A}^s$ in Eq.(\ref{hh2}), moreover, there is another internal U(1) gauge field $\mathbf{A}^a$ between $h$-holon and $a$-spinon in Eq.(\ref{decomp})(cf. the discussion in the paragraph just above Eq.(\ref{heff2})), i.e. $\mathbf{E}_a$ in Eq.(\ref{uniteq2}); in Eq.(\ref{decomp}), the U(1) gauge charges of $a$-spinon and $h$-holon should have opposite sign, and this is the origin of minus sign in Eq.(\ref{uniteq3}).

On the one hand, we have the conservation equation of $j^{\mu}_{s}=(\rho_{\mathrm{spin}},\mathbf{j}_{s})$:
\begin{eqnarray}
 \nabla\cdot\mathbf{j}_{s}+\partial_{t}\rho_{\mathrm{spin}}=0,
\end{eqnarray}
where $[\rho_{\mathrm{spin}}]=[L]^{-2}$. Thus $[\mathbf{j}_{s}]=[L]^{-1}[T]^{-1}$. On the other hand, $\mathbf{E}_{h}=\partial_{t}\mathbf{A}^{h}$ and $
\nabla\times\mathbf{A}^{h}=\pi\hbar\rho_{h}$.  $[\rho_{h}]=[L]^{-2}$, thus $[\mathbf{A}^{h}]=[\hbar][L]^{-1}$ and $[\mathbf{E}_{h}]=[\mathbf{A}^{h}][T]^{-1}=[\hbar][L]^{-1}[T]^{-1}$.

Finally, one gets the unit of $b$-spinon conductivity
\begin{eqnarray}
[\sigma_{s}]=[\mathbf{j}_{s}][\mathbf{E}_{h}]^{-1}=[\hbar]^{-1}
\end{eqnarray}
and similarly, $[\sigma_{h}]=[\sigma_{a}]=[\sigma_{s}]=[\hbar]^{-1}$.

\section{The calculation of the $b$-spinon conductivity $\sigma_s$}
\label{conductivity}

In the LPP, the excited $b$-spinons are deconfined and free, which will decide the longitudinal resistivity via the non-Ioffe-Larking rule (\ref{eq:rhoe}).

The $b$-spinon conductivity $\sigma_s$ can be calculated by the Kubo formula as follows
\begin{eqnarray}
\sigma_s^{\alpha\beta}(\omega)=\frac{i}{\omega}\Pi_s^{\alpha\beta}(\mathbf{q}=0,i\omega_n\rightarrow\omega+i0^{+}).
\end{eqnarray}
Here the polarization tensor in the real space/imaginary time is given by
\begin{eqnarray}
\Pi_s^{\alpha\beta}(i,i';{\tau})\equiv\Pi_{\mathrm{curr.}}^{\alpha\beta}(i,i';{\tau})+\Pi^{\alpha\beta}_{\mathrm{diam.}},
\end{eqnarray}
where $\Pi_{\mathrm{diam.}}$ represents the diamagnetic term of the
polarization tensor, and $\Pi_{\mathrm{curr.}}^{\alpha\beta}(i,i';{\tau})$
denotes the spinon current-current correlation function:
\begin{eqnarray}
\Pi_{\mathrm{curr.}}^{\alpha\beta}(i,i';{\tau})\equiv-\left<T_{{\tau}}J_{s}^{i+\hat{\alpha},i}({\tau})J_{s}^{i'+\hat{\beta},i'}(0)\right>,
\end{eqnarray}
in which the spinon current density $J_{s}^{i+\alpha,i}$ is defined by
\begin{eqnarray}
J_{s}^{i+\hat{\alpha},i}=\left(-iJ_{s}\sum_{\sigma}\sigma e^{i\sigma
\bar{A}^{h}_{i+\hat{\alpha},i}}b^{\dagger}_{i+\hat{\alpha}\sigma}b^{\dagger}_{i\bar{\sigma}}+\text{h.c.}\right)\hat{\alpha}.
\end{eqnarray}

Define the Matsubara Green's function for the Bogoliubov quasiparticle of
the $b$-spinons:
\begin{eqnarray}
G_{\sigma}(m,{\tau})=-\left<T_{{\tau}}\gamma_{m\sigma}({\tau})\gamma^{\dag}_{m\sigma}\right>
\end{eqnarray}
After taking the Fourier transformation, the mean-field solution is given by
\begin{eqnarray}
G^{0}_{\sigma}(m,i\omega_n)=\frac{1}{i\omega_n-E_m},
\end{eqnarray}
where $i\omega_n$ is the bosonic Matsubara frequency $\omega_n=2n{\pi}k_{B}T$.
One may further introduce the spectral function $A_\sigma(m,\omega)$ such that
\begin{eqnarray}
G_{\sigma}(m,i\omega_n)={\int}d\omega\frac{A_\sigma(m,\omega)}{i\omega_n-\omega},
\end{eqnarray}
where
\begin{eqnarray}
A_\sigma(m,\omega)\equiv-\frac{1}{\pi}\mathrm{Im}G_{\sigma}(m,i\omega_n\rightarrow\omega+i0^{+})
\end{eqnarray}
At the mean-field level, the spectral function simply reduces to
$A^0_\sigma(m,\omega)=\delta(\omega-E_{m\sigma})$.
Then the momentum-frequency representation of the spinon current-current correlation can be obtained:
\begin{eqnarray}
\Pi_{\mathrm{curr.}}^{\alpha\beta}(\mathbf{q},i\omega_n)=J^{2}_s\sum_{mm'}F_{mm'}(i\omega_n)G^{\alpha\beta}_{mm'}(\mathbf{q}),
\end{eqnarray}
where
\begin{eqnarray}
G^{\alpha\beta}_{mm'}(\mathbf{q})&=&\frac{1}{N}\sum_{\sigma}\left[\sum_{i}e^{i\mathbf{q}\cdot\mathbf{R}_{i}}e^{i\sigma\bar{A}^{h}_{i+\hat{\alpha},i}}
w^{*}_{m\sigma}(i+\hat{\alpha})w_{m'\sigma}(i)\right]\left[\sum_{i'}e^{-i\mathbf{q}\cdot\mathbf{R}_{i'}}e^{i\sigma\bar{A}^{h}_{i'+\hat{\beta},i'}}
w^{*}_{m'\sigma}(i'+\hat{\beta})w_{m\sigma}(i')\right.\nonumber\\&&
\left.-\sum_{i'}e^{-i\mathbf{q}\cdot\mathbf{R}_{i'}}e^{-i\sigma\bar{A}^{h}_{i'+\hat{\beta},i'}}w_{m\sigma}(i'+\hat{\beta})w^{*}_{m'\sigma}(i')\right],
\label{coefficient}
\end{eqnarray}
Here it is easy to verify that
$G^{\alpha\beta}_{mm'}(\mathbf{q})=\left[G^{\alpha\beta}_{mm'}(-\mathbf{q})\right]^{*}$, and one may define a real number $G^{\alpha\beta}_{mm'}\equiv G^{\alpha\beta}_{mm'}(\mathbf{q}=0)\in\mathbb{R}$. Then
\begin{eqnarray}
F_{mm'}(i\omega_n)&=&\frac{1}{\beta}(u_{m}u_{m'}+v_{m}v_{m'})^2\sum_{i\omega_m}\left[G(m,i\omega_m)G(m',-i\omega_n-i\omega_m)+G(m,i\omega_m)
G(m',i\omega_n-i\omega_m)\right]+\nonumber\\&&\frac{1}{\beta}(u_{m}v_{m'}+u_{m}v_{m'})^2\sum_{i\omega_m}\left[G(m,i\omega_m)G(m',i\omega_n+i\omega_m)
+G(m,i\omega_m)G(m',-i\omega_n+i\omega_m)\right].
\end{eqnarray}

After summing over the Matsubara frequency, we get
\begin{eqnarray}
F_{mm'}(i\omega_n)&=&(u_{m}u_{m'}+v_{m}v_{m'})^2{\int}d{\omega}A(m,\omega){\int}d{\omega'}A(m',\omega')\left[n_B(\omega)-n_B(-\omega')\right]
\left(\frac{1}{i\omega_n+\omega+\omega'}-\frac{1}{i\omega_n-\omega-\omega'}\right)+\nonumber\\&&(u_{m}v_{m'}+u_{m}v_{m'})^2{\int}d{\omega}
A(m,\omega){\int}d{\omega'}A(m',\omega')\left[n_B(\omega')-n_B(\omega)\right]\left(\frac{1}{i\omega_n+\omega-\omega'}-\frac{1}{i\omega_n-\omega+\omega'}\right),
\end{eqnarray}
where
\begin{eqnarray}
A(m,\omega)=\frac{1}{\pi}\frac{\mathrm{Im}\Sigma}{(\omega-E_m-\mathrm{Re}\Sigma)^2+(\mathrm{Im}\Sigma)^2},
\end{eqnarray}
in which $\mathrm{Re}\Sigma$ and $\mathrm{Im}\Sigma$ denote the real part and imaginary part, respectively,
of the self-energy of the $b$-spinon.
%So we can see that if
%we can get the information of the self-energy of spinon, we can
%calculate spin conductivity.

Substituting the mean-field result
$A(m,\omega)=\delta(\omega-E_m)$, one obtains
\begin{eqnarray}
F_{mm'}(i\omega_n)&=&(u_{m}u_{m'}+v_{m}v_{m'})^2\left[n_B(E_m)-n_B(-E_{m'})\right]\left(\frac{1}{i\omega_n+E_m+E_{m'}}-\frac{1}{i\omega_n-E_m-E_{m'}}\right)+\nonumber\\&&
(u_{m}v_{m'}+u_{m}v_{m'})^2\left[n_B(E_{m'})-n_B(E_m)\right]\left(\frac{1}{i\omega_n+E_m-E_{m'}}-\frac{1}{i\omega_n-E_m+E_{m'}}\right).
\end{eqnarray}
And the diamagnetic term of the polarization tensor is given by
\begin{eqnarray}
\Pi^{\alpha\beta}_{MF\mathrm{diam.}}(i-i',{\tau})=2J_s\Delta^s\delta_{\alpha\beta}\delta_{ii'}\delta({\tau}).
\end{eqnarray}
Numerically, we have checked that the diamagnetic term of the polarization tensor gets precisely canceled:
\begin{eqnarray}
\mathrm{Re}\Pi_{\mathrm{curr.}}^{\alpha\alpha}(\mathbf{q}=0,i\omega_n=0)=-2J_s\Delta^s=-\Pi^{\alpha\alpha}_{MF\mathrm{diam.}}(\mathbf{q}=0,i\omega_n=0)
,\label{diacancel}
\end{eqnarray}

Now we consider
\begin{eqnarray}
\mathrm{Re}\sigma_s^{\alpha\beta}(\omega=0)&=&-\frac{\mathrm{Im}{\Pi_s^{\alpha\beta}(\mathbf{q}=0,i\omega_n\rightarrow\omega+i0^{+}})}{\omega}\bigg|_{\omega\rightarrow0}
\nonumber\\&=&-J^{2}_s\sum_{mm'}G^{\alpha\beta}_{mm'}\frac{{\mathrm{Im}}F_{mm'}(i\omega_n\rightarrow\omega+i0^{+})}{\omega}\bigg|_{\omega\rightarrow0}.
\end{eqnarray}
where
\begin{eqnarray}
\frac{\mathrm{Im}F_{mm'}(i\omega_n\rightarrow\omega+i0^{+})}{\omega}\bigg|_{\omega\rightarrow0}&=&2\pi(u_{m}u_{m'}+v_{m}v_{m'})^2{\int}d{\tilde{\omega}}
A(m,\tilde{\omega})A(m',-\tilde{\omega})\frac{{\partial}n_B(\tilde{\omega})}{{\partial}\tilde{\omega}}-\nonumber\\&&2\pi(u_{m}v_{m'}+v_{m}u_{m'})^2
{\int}d{\tilde{\omega}}A(m,\tilde{\omega})A(m',\tilde{\omega})\frac{{\partial}n_B(\tilde{\omega})}{{\partial}\tilde{\omega}}.
\end{eqnarray}

Finally, we arrive at
\begin{eqnarray}
\mathrm{Re}\sigma_s^{\alpha\beta}(\omega=0)=2{\pi}J^{2}_s\sum_{mm'}(u_{m}v_{m'}+v_{m}u_{m'})^2G^{\alpha\beta}_{mm'}{\int}d{\tilde{\omega}}A(m,\tilde{\omega})
A(m',\tilde{\omega})\frac{{\partial}n_B(\tilde{\omega})}{{\partial}\tilde{\omega}},
\end{eqnarray}
Or, after recovering the SI unit, i.e. $[\sigma_{s}]=[\hbar]^{-1}$,
\begin{eqnarray}
\mathrm{Re}\sigma_s^{\alpha\beta}(\omega=0)=\frac{2{\pi}}{\hbar}J^{2}_s\sum_{mm'}(u_{m}v_{m'}+v_{m}u_{m'})^2G^{\alpha\beta}_{mm'}{\int}d{\tilde{\omega}}
A(m,\tilde{\omega})A(m',\tilde{\omega})\frac{{\partial}n_B(\tilde{\omega})}{{\partial}\tilde{\omega}}.
\label{sigmass}
\end{eqnarray}

\section{Derivation of  Eq. (\ref{nonILgeneral})}
\label{gnonIL}

First of all,  we may list the following useful formulas:
   \begin{align}
&   \mathbf{j}_e=\sigma_e \mathbf{E}_e\,\, \label{eq1}\\
  &\mathbf{j}_s=\sigma_s\mathbf{E}_h \,\,\label{eq2}\\
  &\mathbf{j}_h=\sigma_h (\mathbf{E}_s+e\mathbf{E}_e+\mathbf{E}_a) \,\,\label{eq3}\\
    &\mathbf{j}_a=-\sigma_a\mathbf{E}_a \,\,\label{eq4}\\
  &\mathbf{j}_s=\frac{1}{\pi}\epsilon\cdot\mathbf{E}_s\,\,\label{eq5}\\
  &\mathbf{j}_h=\frac{1}{\pi}\epsilon\cdot\mathbf{E}_h\,\,\label{eq6}\\
  &\mathbf{j}_e=e\mathbf{j}_h=e\mathbf{j}_a\,\,\label{eq7}
\end{align}
The notations are defined as follows: $\sigma_{s/h/a}$ denote the `conductivity' of $b$-spinons / $h$-holons / $a$-spinons. $\mathbf{j}_{s/h/a}$ denote their currents. $\epsilon\equiv\left(\begin{smallmatrix}
      0 & 1\\
        -1 & 0
            \end{smallmatrix} \right)$ is a matrix acting on $\hat{x}$- and $\hat{y}$- coordinates, $\epsilon^{12}=\epsilon^{xy}=1,\epsilon^{21}=\epsilon^{yx}=-1$. $\mathbf{j}_e$ is the electric current that is physically detected in transport experiments. $\mathbf{E}_{s/h/a}=-\partial_t\mathbf{A}^{s/h/a}$ are electric fields formed by gauge fields $\mathbf{A}^{s/h/a}$.

The physical pictures of the seven formulas are explained as follows. Eq. (\ref{eq1}) is the definition of the well-known Ohm's law.  $\mathbf{A}^e$ is external electromagnetic field. Eq. (\ref{eq2}) is a response formula in analog to `Ohm's law', meaning that `electric field' $\mathbf{E}_h$ formed by $\mathbf{A}^h$ generates a $b$-spinon current owing to the minimal coupling between $\mathbf{A}^h$ and $b$-spinons. This minimal coupling can be found in  Eqs. (\ref{b-RVBpairing}) and (\ref{hs2}). Likewise,  Eq.(\ref{eq3}) is a response formula about $h$-holon current.  $h$-holons simultaneously couple to three gauge fields, namely, $\mathbf{A}^s$ (i.e. Eq. (\ref{hh2})), $\mathbf{A}^e$ (i.e.  Eq. (\ref{hh2})), and, $\mathbf{A}^a$ (see the discussion above Eq. (\ref{heff2})).  Due to opposite $\mathbf{A}^a$ gauge charges carried by $h$-holons and $a$-spinons, Eq. (\ref{eq4}) that describe the linear response of $a$-spinons can also be easily understood.

 Eqs. (\ref{eq5}) and (\ref{eq6}) can be understood via  Eqs. (\ref{cond1}) and (\ref{cond2}). For instance, Eq. (\ref{cond2}) indicate that $h$-holon particle density is the source of the magnetic flux of $\mathbf{A}^h$ gauge field. Therefore, once $h$-holons moves and thereby there is a holon current $\mathbf{j}^h$,  $h$-holons will necessarily generate electric field of $\mathbf{A}^h$ along the transverse direction. More rigorous derivation of   Eqs. (\ref{eq5}) and (\ref{eq6}) can be performed in the mutual Chern-Simons gauge field theory which has space-time covariant form as shown in Ref. \onlinecite{YTQW_11}.

 The first identity in Eq.  (\ref{eq7}) is obvious since  each $h$-holon carries a fundamental electric charge while $a$-spinons and $b$-spinons are charge-neutral in our fractionalization framework. The second identity in Eq. (\ref{eq7}) can be understood as a consequence of the internal gauge field $\mathbf{A}^a$. More pictorially, $h$-holons and $a$-spinons are created and annihilated together implied by Eq. (\ref{decomp}). In the following, we may apply these seven formulas to derive Eq. (\ref{nonILgeneral}).

$\mathbf{E}_s$ in Eq. \eqref{eq5} can be expressed as: $\mathbf{E}_s=\epsilon^{-1}\cdot\mathbf{j}_s\pi=-\epsilon\cdot \mathbf{j}_s\pi$. $\mathbf{E}_h$ in Eq. \eqref{eq6} can be expressed as: $\mathbf{E}_h=\epsilon^{-1}\cdot\mathbf{j}_h\pi=-\epsilon\cdot\mathbf{j}_h\pi$. Further consider \eqref{eq2}  , we end up with:
\begin{align}
\mathbf{E}_s=-\epsilon\cdot\mathbf{j}_s\pi=-\epsilon\cdot\mathbf{E}_h\sigma_s\pi=\epsilon\cdot\epsilon\cdot \mathbf{j}_h\sigma_s\pi^2=-\mathbf{j}_h\sigma_s\pi^2
\end{align}
Consider Eq. (\ref{eq7}) and Eq. (\ref{eq1}), we have:
\begin{align}
\mathbf{E}_s=-\mathbf{j}_h\sigma_s\pi^2=-\mathbf{j}_e\frac{\sigma_s\pi^2}{e}
=-\mathbf{E}_e\frac{\sigma_e\sigma_s\pi^2}{e}
\label{eqes}
\end{align}
Consider Eqs. (\ref{eq7}), (\ref{eq1}) and (\ref{eq4}), we have:
\begin{align}
\mathbf{E}_a=-\sigma^{-1}_a\mathbf{j}_a=-\frac{\sigma^{-1}_a}{e}\mathbf{j}_e
=-\frac{\sigma^{-1}_a\sigma_e}{e}\mathbf{E}_e\label{eqea}
\end{align}
Substituting (\ref{eqes}) and (\ref{eqea}) into Eq. (\ref{eq3}), we end up with:
\begin{align}
\mathbf{j}^h=\sigma_h (e\mathbf{E}_e-\frac{\sigma_e\sigma_s\pi^2}{e}\mathbf{E}_e-\frac{\sigma_a^{-1}\sigma_e}{e}\mathbf{E}_e)
\end{align}
By further considering $e\mathbf{j}_h=\mathbf{j}_e=\sigma_e\mathbf{E}_e$, for any $\mathbf{E}_e$, the following identity is valid:
\begin{align}
\sigma_e=\sigma_h (e^2-\sigma_e\sigma_s\pi^2-\sigma_a^{-1}\sigma_e)
\end{align}
which is identical to:
\begin{align}
\sigma^{-1}_e=\frac{1}{e^2}\left(\sigma^{-1}_h  +\sigma_a^{-1}+\pi^2\sigma_s\right)
\end{align}
Once the SI unit is recovered, Eq. (\ref{nonILgeneral}) is obtained:
\begin{align}
\sigma^{-1}_e=\frac{1}{e^2}\left(\sigma^{-1}_h  +\sigma_a^{-1}+\pi^2\hbar^2\sigma_s\right)
\end{align}

\end{widetext}


\begin{thebibliography}{99}
\bibitem{comment1} Note that here the expression of  $T_0$ in Eq. (\ref{T0}) is slightly different from Eq. (15) given in Ref. \onlinecite{GZC2005} because in the latter a slightly different constraint for $b$-spinons, i.e., $\sum_{i\sigma}b^{\dag}_{i\sigma}b_{i\sigma}=N(1-\delta)$, is used in the mean-field theory.
\end{thebibliography}
\end{document}